\documentclass[format=acmsmall, review=false, screen=true]{acmart}

\usepackage{booktabs} 
\usepackage[position=b]{subcaption}
\usepackage{amsmath,amssymb}
\usepackage{bbm}
\usepackage{tabularx}
\DeclareMathOperator*{\argmax}{arg\,max}
\DeclareMathOperator*{\argmin}{arg\,min}
\graphicspath{{images/},
{images/partial-data_GAN/}
}
\usepackage[ruled]{algorithm2e} 

\acmJournal{TOMM}
\acmVolume{9}
\acmNumber{4}
\acmArticle{39}
\acmYear{2018}
\acmMonth{11}

\setcopyright{acmcopyright}

\acmDOI{0000001.0000001}

\begin{document}
\title{Synthesizing facial photometries and corresponding geometries 
     using generative adversarial networks }
 
\author{Gil Shamai*}
\orcid{1234-5678-9012-3456}
\affiliation{%
  \institution{Technion -- Israel Institute of Technology}
  \city{Haifa}
  \country{Israel}
  }
\email{gil.shamai@gmail.com}
\author{Ron Slossberg*}
\affiliation{%
  \institution{Technion -- Israel Institute of Technology}
  \city{Haifa}
  \country{Israel}
}
\email{ronslos@gmail.com}
\author{Ron Kimmel}
\affiliation{%
  \institution{Technion -- Israel Institute of Technology}
  \city{Haifa}
  \country{Israel}
 }
\email{ron.kimmel@gmail.com}
* Indicates equal contribution by authors

\begin{abstract}
Artificial data synthesis is currently a well studied topic with useful applications in data science, computer vision, graphics and many other fields. Generating realistic data is especially challenging since human perception is highly sensitive to non-realistic appearance. In recent times, new levels of realism have been achieved by advances in GAN training procedures and architectures. These successful models, however, are tuned mostly for use with regularly sampled data such as images, audio and video. Despite the successful application of the architecture on these types of media, applying the same tools to geometric data poses a far greater challenge. The study of geometric deep learning is still a debated issue within the academic community as the lack of intrinsic parametrization inherent to geometric objects prohibits the direct use of convolutional filters, a main building block of today's machine learning systems. 

In this paper we propose a new method for generating realistic human facial geometries coupled with overlayed textures. We circumvent the parametrization issue by imposing a global mapping from our data to the unit rectangle. This mapping enables the representation of our geometric data as regularly sampled 2D images. We further discuss how to design such a mapping to control the mapping distortion and conserve area within the mapped image. By representing geometric textures and geometries as images, we are able to use advanced GAN methodologies to generate new geometries. We address the often neglected topic of relation between texture and geometry and propose to use this correlation to match between generated textures and their corresponding geometries. In addition, we widen the scope of our discussion and offer a new method for training GAN models on partially corrupted data. Finally, we provide empirical evidence demonstrating our generative model’s is ability to produce examples of new identities independent from the training data while maintaining a high level of realism, two traits that are often at odds.

\end{abstract}

%
%


%
%


\maketitle

\section{Introduction}
The generation of realistic examples of everyday objects is a challenging and interesting problem which relates to several research fields such as geometry, computer graphics, and computer vision. The ability to capture the essence of a class of objects is key to the task of generating diverse datasets which may be used in turn during the training of many machine learning based algorithms. The main challenge posed by the task of data generation is to construct the model that is able to generalize to many variations while still maintaining high detail and quality. Furthermore, the challenge of generating geometric data is even greater since Both geometry and texture of an object must be synthesized while taking into account the underlying relations between them.

In this work, we propose to learn the latent space of 3D textured objects. We focus our efforts on human faces, and show that by using a canonical transformation that maps geometric data to images, we are able to learn the distribution of such images via the GAN framework. By representing both texture and geometry of the face as transformed geometric images, we can learn the underlying distribution of faces, and later generate new faces at will. The generation of realistic human faces is a useful tool with applications in face recognition, puppetry, reconstruction and rendering. Our main contributions are the proposition of a new model for 3D human faces which is composed in the 2D image domain, as well as the modeling of the relation between texture and geometry, further improving realism. By generating geometries and textures using state of the art GANs, it is possible create highly detailed data samples while maintaining the ability go generalize to unseen data, two desirable propertied that are often at odds. 

While deep learning and convolutional networks have revolutionized many fields in recent years, they have been mostly employed on structured data which is intrinsically ordered. Arranged data such as audio, video, images, and text can be processed according to the order of samples, frames, pixels or words. This inherent ordering permits the application of convolution operations which are the main building block of convolutional networks, a powerful and popular variant of deep networks. Contrary to typical parameterized data, geometric data represented by two dimensional manifolds lacks an intrinsic parameterization and is therefore more difficult to process via convolutional networks. This important class of data is crucial to the task of modeling our world as most solid objects can be represented by a closed manifold accompanied by a texture overlay. 

Recently, geometric data has grown dramatically in availability as more accurate and affordable acquisition devices have come into use. This abundance of data has attracted the attention of the computer vision and machine learning communities, leading to many new approaches for modeling and processing of geometries. One family of techniques for geometric data processing aims to define new operators which can be applied directly to the manifold and are able to replace to some extent the convolution operation within the processing pipeline. Other methods attempt to process geometries in the spectral domain or represent them in voxel space. These families of methods each have their merits but suffer from other issues such as loss of generality and memory inefficiency. In contrast, we propose to transform our geometric data via a canonical mapping into two dimensional gridded data. This allows us to process the geometric data as images. While this approach on its own is not new we show that by careful construction of the transformed dataset we are able to harness the power of convolutional networks with little loss of data fidelity. Furthermore, we are able to design our transformation process in order to control the distortion, thus reducing it in important areas while spreading it to the non essential areas of the data. Finally, we propose to encode both the geometry and texture as mapped images which means the processing pipeline remains identical for both cases.

\section{Related work}
Data augmentation is a common practice within the machine learning community. By applying various transformations to existing data samples it is possible to simulate a much larger dataset than is available and introduce robustness to transformations. A more advanced method for data augmentation takes into account the geometry of the scene. The technique which we term geometric data augmentation consists of a geometry recovery stage, then transformation is performed on the geometry and finally a new image is created by projecting the geometry. In \cite{masi2016we}, the authors show that by performing geometric data augmentation on a dataset of facial images they are able to reach state of the art results on difficult facial recognition benchmarks. Despite its proven usefulness geometric augmentation still lacks the ability to create completely new data samples outside the scope of the dataset.

A complementary method to data augmentation is data generation. Bu constructing a high quality model for data generation it is possible to produce an infinitely large dataset. In addition, some models may permit control over the characteristics of each data sample. Within the domain of faces this would mean control over parameters such as age, gender, expression, pose and lighting conditions. When dealing with image data a recent popular approach is to use a GAN \cite{goodfellow2014generative} which is in essence a neural network with a trainable loss function. While this class of methods is well suited for images, reformulation in the context of geometry is more challenging and several competing approaches exist in this field. \cite{Gecer_2018_ECCV} and \cite{shrivastava2017learning} propose to construct samples from a low quality linear model, and then use a GAN in order enforce the realism of the data. \cite{Litany_2018_CVPR} and \cite{ranjan2018generating} both propose the use of convolutional autoencoders which are trained on pre-aligned geometric data. These methods however do not take into account the model texture. In addition \cite{wu2016learning} have used the popular voxel grid representation for geometries, and are able to generate 3D objects using this notion. This method however is memory inefficient and in practice can produce only coarse geometries. 

In addition to data augmentation and generation the objective of pose normalization is to decouple the subjects identity from other factors such as expression and pose which may confuse a classifier. This can be either done by geometric reconstruction manipulation of the facial geometry or by performing normalization directly in the image domain. While \cite{chu20143d} and \cite{bas20173d} leverage a geometric representation in order to transform the data, \cite{tran2017disentangled} and \cite{huang2017beyond} are able to frontalize faces directly in the image domain as part of their pipeline. Although useful methods which help the training process by limiting data variation, these methods still do not explicitly model new data samples which is our ultimate goal. 

An additional method for geometrically manipulating facial data which has gained success is geometric reconstruction from a single image. One popular family of methods aim to fit a parametric model to an image. This idea was first introduced by \cite{blanz1999morphable} and has since been extended by works such as \cite{booth20173d}. An approach which involves regressing the coefficients of a given model via a deep network were popularized by \cite{richardson20163d} and extended by \cite{richardson2017learning} and \cite{tran2017regressing}. More recently methods which are not restricted to a specific model or attempt to learn the model during training time such as \cite{sela2017unrestricted}, \cite{Tewari_2018_CVPR} and \cite{tran2018nonlinear} have been able leave the restricting assumptions of linear models such as 3DMM. Complementary efforts such as 
\cite{Deng_2018_CVPR} propose to reconstruct occluded texture regions in order to gain a full textured reconstruction from challenging poses as well. Another recent work by \cite{saito2017photorealistic} focuses on improving the quality of facial texture used in reconstructed faces in order to improve realism. An additional complimentary approach proposed by \cite{Guler2016DenseReg} is to learn a direct mapping from an image to a template model. All of the above approaches while useful, are based on fitting some geometry to a given image by relying on some underlying geometric model. This model however is not explicitly used in order to generate novel faces but rather to reconstruct existing ones.

Our most direct competition comes from several works in the field of facial generative modeling. The seminal work by \cite{blanz1999morphable} which pioneered the field almost two decades ago is still widely used within many methods, some of which were mentioned above. The linear 3D Morphable Model proposed is extremely flexible; however it has the drawback of using a small number of PCA vectors which limit its ability to present highly detailed models. A recent large scale effort taken by \cite{booth20163d} and \cite{booth2018large} has produced the largest publicly known 3DMM by scanning $10k$ subjects and using their scans to construct the model. In contrast to linear models much more complex relations can be captured by training deep networks to take the part of data generators. To this end, \cite{Tewari_2018_CVPR} and \cite{tran2018nonlinear} were able to jointly learn a reconstruction encoder while also learning the facial model itself. Given the trained model one could plausibly generate faces, however the authors have not shown any experiments to this effect.  \cite{ranjan2018generating} on the other hand has employed mesh autoencoders to construct new facial geometries, however this method does not produce texture and was trained on a limited dataset of very few subjects. In this work we will propose a new GAN based facial geometric generative model, and analyze the ability of our model to extend to new identities. We also relate between the geometric and texture models which are intrinsically correlated and discuss different ways of exploiting this correlation for our cause.

\section{3D Morphable Model}
\label{sec:3dmm}
One of the early attempts to capture facial geometry and photometry (texture) by a linear low dimensional space is the Blanz and Vetter \cite{blanz1999morphable} {\em 3D Morphable Model} (3DMM). 
Using the 3DMM, textures and geometries of faces can be synthesized as a linear combination of the elements of an orthogonal basis. 
The basis is constructed from a collection of facial scans and by applying the principal component analysis after alignment of the faces.  
That is, the basis construction process relies on a vertex to vertex alignment of the facial scans, which is achieved by computationally finding a dense correspondence between each scan to a template model. 
The aligned vertices provide a set of spatial and texture coordinates which are then decomposed into the principal components of the set. 
Once the basis is constructed, it is possible to represent each face by projecting it onto the first $k$ components of both the geometry and the texture bases.

This linear model was used to reconstruct 3D faces from 2D images; 
Blanz and Vetter \cite{blanz1999morphable} took an analysis-by-synthesis approach, which attempts to fit a projected surface model embedded in $\mathbb{R}^3$ into a given 2D image. 
This was established by constructing a fully differentiable parametric image formation pipeline, and performing a gradient descent procedure optimizing for an image to image loss on the model parameters. 
The parameters consist of the geometry and texture models coefficients of the face, as well as the lighting and pose parameters. 
This process results in a set of coefficients which encode the geometry and texture of any given face up to their projections on the principal components basis, effectively reconstructing the curved surface structure and the photometry of the given image of a face.

\subsection{Model Construction}
According to the 3DMM model, each face is represented as an ordered set of $m$ geometric coordinates \mbox{$g = (\hat x^1, \hat y^1, \hat z^1, \hat x^2,\ldots,\hat y^m, \hat z^m) \in \mathbb{R}^{3m}$}  and texture coordinates in RGB space\\ \mbox{$t = (\hat r^1, \hat g^1, \hat b^1, \hat r^2,\ldots,\hat g^m, \hat b^m) \in \mathbb{R}^{3m}$}.
Given a set of $n$ faces, each represented by geometry $g_i$ and texture $t_i$ vectors, construct the $3m \times n$ matrices $G$ and $T$ by column wise concatenation of all geometric coordinates and all corresponding texture coordinates.
Since the alignment process ensures an ordered universal representation of all faces, Principal Component Analysis (PCA) \cite{jolliffe1986principal} can be applied to extract the optimal first $k$ orthogonal basis components in terms of $L_2$ reconstruction error.
To that end, denote by $V_g$ and $V_t$ the $3m \times n$ matrices that contain the left singular vectors of $\Delta G = G - \mu_g\mathbbm{1}^T$ and $\Delta T = T - \mu_t\mathbbm{1}^T$, respectively, where $\mu_g$ and $\mu_t$ are the average geometry and texture of the faces and $\mathbbm{1}$ is a vector of ones. 

By ordering $V_g$ and $V_t$ according to the magnitude of the singular values in a descending order,
the texture and the geometric coordinates of each given face can be approximated by the linear combination
\begin{equation}
g_i = \mu_g + V_g\alpha_{g_i},\ \ \  t_i = \mu_t + V_t\alpha_{t_i},
\label{eq:3dmm_model}
\end{equation}
where $\alpha_{g_i}$ and $\alpha_{t_i}$ are the coefficients vectors,  obtained by $\alpha_{g_i} = V_g^T(g_i - \mu_g)$ and $\alpha_{t_i} = V_t^T(t_i - \mu_t)$.
Following this formulation, it is possible to use such a model to generate new faces by randomly selecting the geometry and texture coefficients and plugging them into \autoref{eq:3dmm_model}.
According to \cite{blanz1999morphable}, the distribution of the coefficients can be approximated as a multivariate normal distribution, such that the probability for a coefficient vector $\alpha$ is given by
\begin{equation}
P(\alpha) \sim \exp\left \{-\frac{1}{2}\alpha^T\Sigma^{-1}\alpha\right \},
\label{eq:3dmm_dist}
\end{equation}
where $\Sigma$ is a covariance matrix that can be empirically estimated from the data, and is generally assumed to be diagonal.

\subsection{Synthesis model}
The 3D morphable model is useful not only in the context of representation and reconstruction, but, as noted in the previous section, it also allows for the generation of new faces which can not be found in the training set. 
The synthesis is achieved by randomizing linear combinations of the basis vectors. 
The random coefficients are drawn according to the model prior from the distribution described in \autoref{eq:3dmm_dist}. 
As is common practice when dealing with principal components, only the first $k \ll n$ 
vectors can be taken into account as part of the model. 
The number $k$ can be obtained by analyzing the decay of the singular values which is proportional to the error produced by ignoring the associated basis vector. 
By excluding the vectors for which the singular variables are sufficiently small we can guarantee minimal loss of data. 

\begin{figure}
\centering
\includegraphics[width=0.9\linewidth]{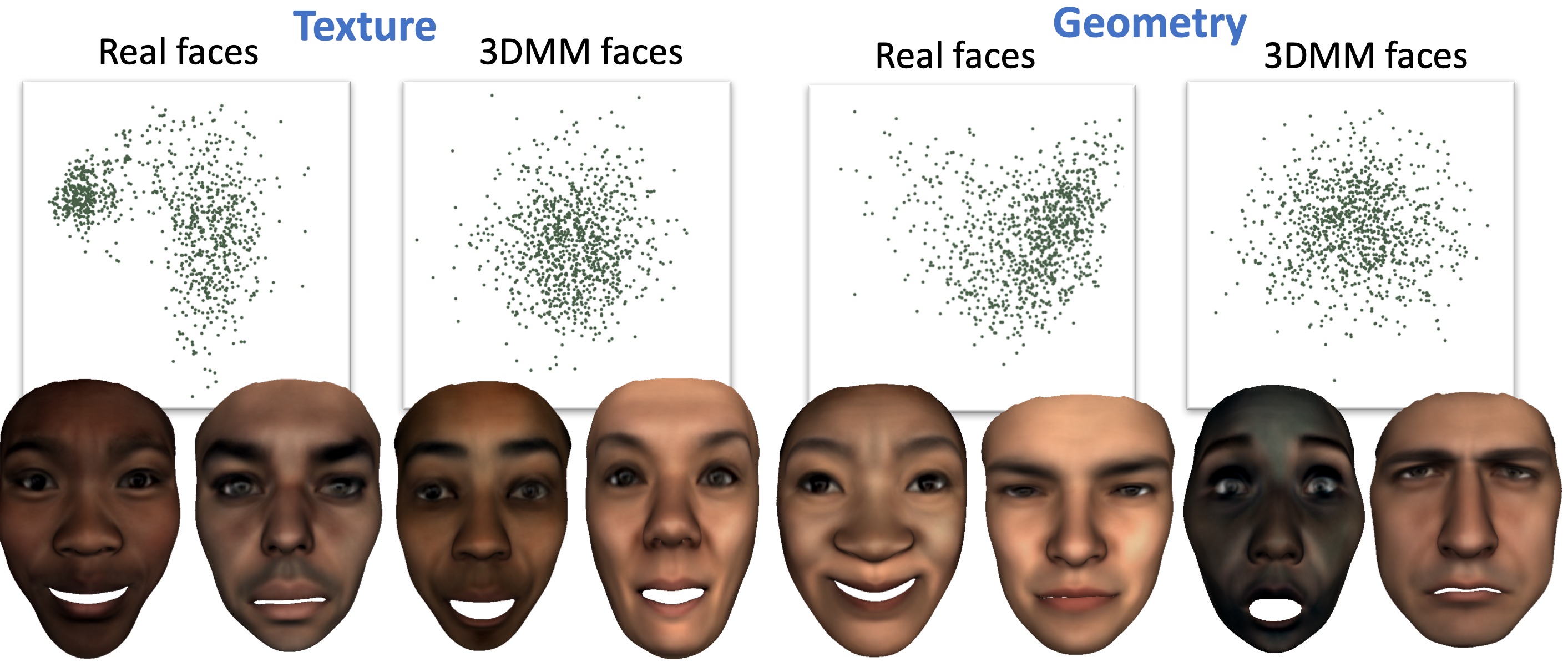}
\caption{Top: Multi Dimensional Scaling \cite{shamai2018efficient} used to depict the first $k=200$ 3DMM texture and geometry coefficients for real faces versus 3DMM generated faces. Bottom: Examples of 3DMM generated faces. The 3DMM was constructed by our own training set. }
\label{fig:3DMM_faces}
\end{figure}

Even though two decades have passed since the inception of the 3DMM, it is still widely used in cutting edge applications. 
By harnessing the generative powers of this model, it has been used as a tool for data augmentation and data creation for training of convolutional networks \cite{sela2017unrestricted,richardson20163d,richardson2017learning,Gecer_2018_ECCV}. 
Furthermore, the model has been integrated into deep learning pipelines in order to provide structure and regularization to the learning process \cite{Tewari_2018_CVPR}. 
In spite of the wide use and apparent success of the model it is clear that the faces obtained from it tend to be over-smoothed and in some cases non-realistic. 
Furthermore, the multivariate normal distribution model from which the coefficients are drawn is over simplified and does not represent the true distribution of faces.
In particular, the texture and geometry are treated as two uncorrelated variables, in contradiction to empirical evidence. \autoref{fig:3DMM_faces} shows a few samples of synthesized 3DMM faces and depicts the difference between the distributions of 3DMM generated faces and real ones.

\section{Progressive growing GAN}
\label{sec:GAN}
Generation of novel plausible data samples requires learning the underlying distribution of the data. Given a perfect discriminator which can differentiate between real and fake data samples it is possible to construct a training loss for a generator model which tries to maximally confuse the discriminator. 
For complex realistic data, finding such a discriminator is a difficult problem on its own and requires learning from realistic and fake examples. 

The fundamental idea of the GAN framework is to train both of these networks simultaneously. 
Essentially, this means that we use a trainable loss function for the generator which constantly evolves as the generator improves. 
This process can be formulated as \autoref{eq:minimaxgame-definition}
\begin{equation}
\footnotesize
\label{eq:minimaxgame-definition}
\min_G \max_D V(D, G) = \mathbb{E}_{x \sim p_{\text{data}}(x)}[\log D(x)] + \mathbb{E}_{z \sim p_{z}(z)}[\log (1 - D(G(z)))],
\end{equation}
where $D,G$ are the discriminator and generator parametric functions, $x,z$ are the real data samples and latent representation vector respectively.

Since we wish to produce high resolution textures for facial geometries, we propose to use a recent successful GAN, namely \cite{karras2017progressive}. 
The progressive growing GAN is built in levels which gradually increasing the resolution of the output image. 
During the training process each level is added consecutively while smoothly blending the new levels into the output as they are added. 
This, and several other techniques were shown to increase the training stability as well as the variation of the produced data.

The difficulty concerning geometric data is that it lacks the regular intrinsic ordering which exists in 2D images, which are essentially large matrices. 
For this reason, it is unclear how to apply spatial filtering, which is the core building block of neural network layers, to arbitrary geometric structures. 
Significant progress has been made in this direction by several recent papers. A comprehensive survey is presented in \cite{bronstein2017geometric}. These methods, however, are not yet widely used and supported within standard deep learning coding libraries.
In order to harness the full power of recent state of the art developments in the field, it is sometimes preferable to work in the domain of images. For this reason, we built a data processing pipeline which maps the geometric scanned data into a flat canonical image which allows the utilization of the progressively growing GAN without major modifications.

\section{Training data construction}
\label{sec:training_data_construction}
In this section we describe the process by which we produce our training data. 
We start with digital geometric scans of human faces. 
By making use of a surface to surface alignment process \cite{weise2009face}, we are able to bring all the scans into correspondence with each other. 
Next, applying a universal mapping from the mesh to the 2D plane, we can transfer the facial texture into a canonically parametrized image. 
These aligned texture images are used to train our texture generation model. 

We provide several alternatives for constructing the facial geometry which accompanies each texture. One solution is to learn the relation between 3DMM texture and geometry coefficients which is prevalent in the training data. 
In addition, we can similarly process the geometric data of the faces as well. By applying the same canonical transformation and encoding the $\left(x,y,z\right)$ coordinates of the model vertices as RGB channels of an image, we can learn to generate geometries as well as textures using the same methodology.

 \begin{figure}
     \centering
     \includegraphics[width=0.8\linewidth, trim={2cm 5cm 2cm 5cm}, clip]{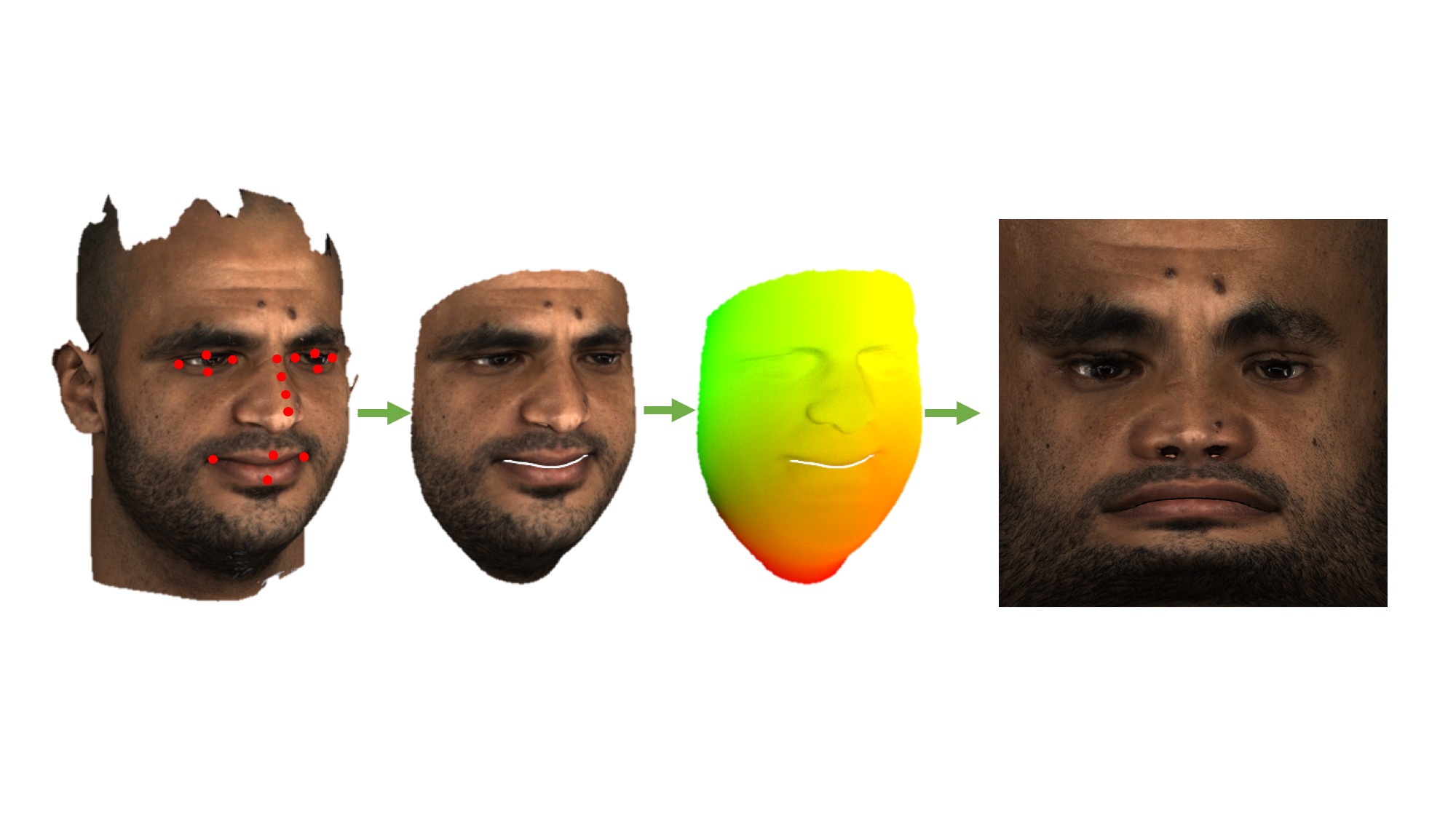}
     \caption{Left to right: Raw scan with landmark points, The template is deformed to fit the scan and texture is transferred onto the template, canonical UV mapping between the deformed template and the 2D image, final mapped texture as image.}
     \label{fig:my_label}
 \end{figure}
\subsection{Face scanning and marking}
Our training data formulation process starts by acquiring digital high resolution facial scans. 
Using a 3DMD scanner, roughly $1000$ different subjects were scanned, each making five distinct facial expressions including a neutral expression. 
The subjects were selected to form a balanced mixture of genders and ethnic backgrounds. 
Each scan is comprised of the facial geometry, represented by a triangulated mesh, as well as two high resolution photographs, which capture a 180 degree view of the subject's face. 
Each mesh triangle is automatically mapped to one of the photos, allowing the facial texture to be transferred onto the mesh. 

Due to the variety of facial geometries, as well as limitations of the scanning process, the meshes may contain imperfections such as holes and areas of missing texture. 
These data corruptions may affect the training data samples that are given to the network and care must be taken not to hinder the training process.
The straightforward path is to filter out the erroneous data samples completely. 
This leads to a significant reduction in the overall size of the training set size of roughly 20\%. Instead, we propose a new approach which incorporates corrupted scans without compromising the integrity of the training data. 
We describe our approach to learning from corrupted data in \autoref{sec:corrupted_data}.

In order to facilitate the alignment process described in \autoref{sec:Non-Rigid_Alignment}, we annotate each face by 43 landmark locations. 
These locations are determined automatically by projecting the facial surface into an image and applying one of many 2D facial landmark detectors such as Dlib \cite{dlib09}. 
The landmarks are then back-projected onto the surface to determine their location. 
Finally, the locations of the automatically generated landmarks are manually refined in order to prevent displacements that could lead to large errors during the alignment process.

\subsection{Non-Rigid Alignment}
\label{sec:Non-Rigid_Alignment}
The goal of the alignment process is to find a dense correspondence between all the facial geometries. 
It is performed by aligning all scans to a single facial template. 
This correspondence is achieved by deforming the template into the scanned surface, a process guided by the pre-computed landmarks. 

Initially, a rigid alignment between the scanned surface and template is performed as preprocessing step. 
This is done by solving for the rotation, translation, and uniform scaling between the scan and template landmarks. 
The deformation process is performed by defining a fitting energy which takes into account both surfaces and known landmarks and measures how closely they fit each other. 
The energy also includes a regularization term which penalizes non-smooth deformations. 
The template mesh is deformed by moving each vertex according to the energy gradient in an iterative manner. 

The loss function which is minimized during the alignment process was first described by \cite{blanz1999morphable} and is comprised of $3$ terms which contribute to the final alignment. 
The first term accumulates the distances between the facial landmark points on the scanned facial surface and their corresponding points on the template mesh.
The second term accumulates the distances between all the template mesh points to the scanned surface. 
The third term serves as a regularization, and penalizes non-smooth deformations. 
The loss term is minimized by taking the derivative of the loss with respect to the template vertex coordinates, then deforming the template in the gradient direction. 
This process naturally provides a dense point to point correspondence between each and every scanned surface.
\subsection{Universal mapping} 
\label{sec:mapping}
Given a facial scanned surface  with unknown parametrization, our goal in this section is to discover a 2D parameterization of the surface which maps it to a unit rectangle, such that this mapping is consistent for all scans.
In \autoref{sec:Non-Rigid_Alignment}, we described the process of aligning the facial surface template to a scanned facial surface, and by that, bring them into correspondence. 
The obtained correspondence allows to transfer the parametrization from the template to all scans, thus establishing a universal parametrization. In the following section, we define the unique parameterization between the template face and the unit rectangle.

\begin{figure}
\centering
\includegraphics[width=0.9\linewidth]{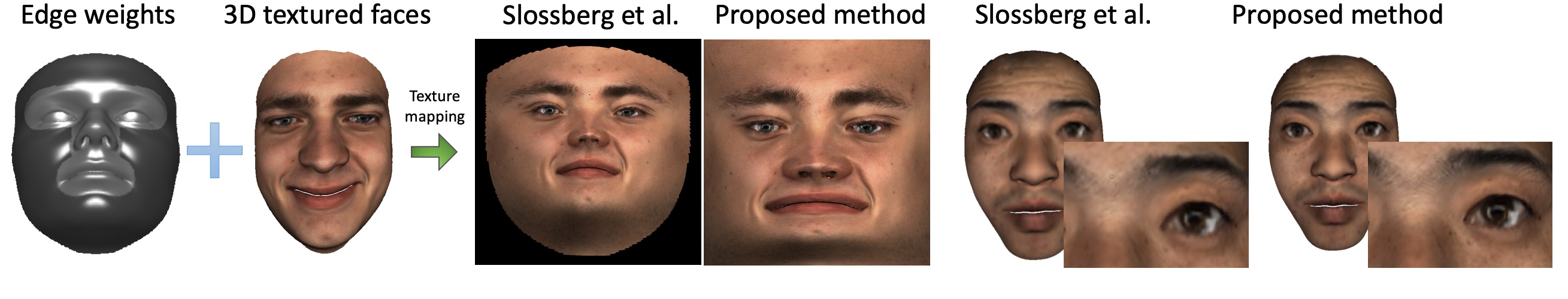}
\caption{Left: A visualization of the proposed edge weights. Each vertex is colored by the average weights of its edges. Bright colors represent larger weights.
Next: The deformed template, after its alignment process to some arbitrary scan. Center: the resulting 2D image, using the texture mapping suggested in \cite{slossberg2018high} and the proposed one. 
Right: A comparison between the texture mapping methods. An unwrapped texture is mapped back to the template, showing a slightly better resolution in the proposed method.}
\label{fig:uv_result}
\end{figure}

The authors of \cite{slossberg2018high} defined tbe mapping between the scan and the plane by using a ray casting technique built into the animation rendering toolbox of Blender \cite{blender}. 
\autoref{fig:uv_result} depicts several examples of the resulting mapped facial photometry.
Although it would be possible to make use of the same parametrizatoin, an alternative definition may suite us better. 
The Blender mapping, for example, does not exploit the entire squared image for the mapping.
Moreover, it does not take the facial structure into account. 
The eyes, nose, and mouth, for instance, clearly  contain more details than smoother parts of the face such as the cheeks and forehead. 
It is reasonable to assume that it would be easier to learn and reconstruct the main features, perhaps at the expense of other parts if they take up a larger portion of the input images. To that end, we propose to construct a weighted parametrization that will allow us to control the relative area in the plane taken up by each facial feature.

In \cite{floater1997parametrization}, the authors presented a parametrization technique that allows to choose for each vertex its baricentric coordinates with respect to its neighbors.
The authors demonstrate that any set of baricentric coordinates has a unique planar graph with a valid triangulation that fulfills it. 
As an extension, they also provided a method for a weighted least square parametrization that allows some control over the edge lengths in the resulting parametrization.
The method is briefly described as below. 

Given any triangulated mesh, the object is to map it into a valid planar graph with the same connectivity. 
Assuming a mesh with $N$ vertices, choose a set of $K$ boundary vertices from the mesh and fix their 2D mapping values to some desired convex boundary, $u_1,...,u_K$.
For any other vertex $i > K$ in the mesh, choose a set of non-negative baricentric coordinates $\lambda_{i,j}$, such that 
$\sum_{j=1}^N \lambda_{i,j} = 1$, and $\lambda_{i,j} = 0$ if and only if $i$ and $j$ are not connected. 
Then, for $i=K+1,...,N$, solve the linear system of equations
\begin{equation}
\label{eq:baricentric}
    u_i = \sum_{j=1}^N \lambda_{i,j}u_j.
\end{equation}
The authors in \cite{floater1997parametrization} prove that \autoref{eq:baricentric} has a closed form unique solution that coincides with the chosen baricentric coordinates. 
According to \cite{floater1997parametrization}, this technique could be extended to a weighted least square parametrization. 
For any desired set of weights $w_{i,j}$, it was shown that the choice of
\begin{equation}
\lambda_{i,j} = \frac{w_{i,j}}{\sum_{j:(i,j)\in E} w_{i,j}},
\end{equation}
minimizes the functional $\sum_{j:(i,j)\in E} w_{i,j} \|u_i - u_j\|^2$, where $E$ represents the set of edges.

Following this technique, we designed the weights $w_{i,j}$ such that eyes, nose and mouth would recieve a larger area in the parametrization plane. We defined a weight for each vertex in the template face, and then gave each edge the average weight of its two adjacent vertices. 
Note that the resulting edge lengths also depend on the density of vertices in the mesh. 
In other words, when choosing a constant weight for all edges, the edge lengths of the resulting parametrization termed the uniform baricentric parametrization, is not constant. 
To design the edge weights more intuitively, we normalize the edge weights by the ones resulting from the uniform baricentric parametrization. 
A visualization of the edge weights is shown in \autoref{fig:uv_result}.

To choose the boundary vertices $u_1,\ldots,u_K$, we follow the outer boundary of the facial mesh, starting from the center bottom (a point on the chin), while measuring the length of edges we pass through, $L_1,...,L_k$. 
Assume the image boundary is parametrized by $C(t) = \{x(t), y(t)\}$ for $0 \leq t \leq 1$, such that $C(0)=C(1)$ is the bottom center of the image. 
Then, we set $u_i = C(t_i)$, where
\begin{equation}
   t_i = \frac{\sum_{j=1}^i L_j}{\sum_{j=1}^K L_j}.
\end{equation}
Lastly, unlike \cite{slossberg2018high}, we propose to construct a symmetric mapping, in order to augment the data by mirroring the training samples.
This could be done by ensuring that the template is intrinsically symmetric, as well as the choice of boundary vertices and edge weight. 
The resulting mapping and a visualization of the edge weights are shown in \autoref{fig:uv_result}.
The rightmost part in \autoref{fig:uv_result} shows that when mapping back the unwrapped texture to the facial geometry, a better resolution is obtained when using the proposed method. 

\section{Learning from corrupted data}  
\label{sec:corrupted_data}
\begin{figure}
\centering
\includegraphics[width=0.8\linewidth]{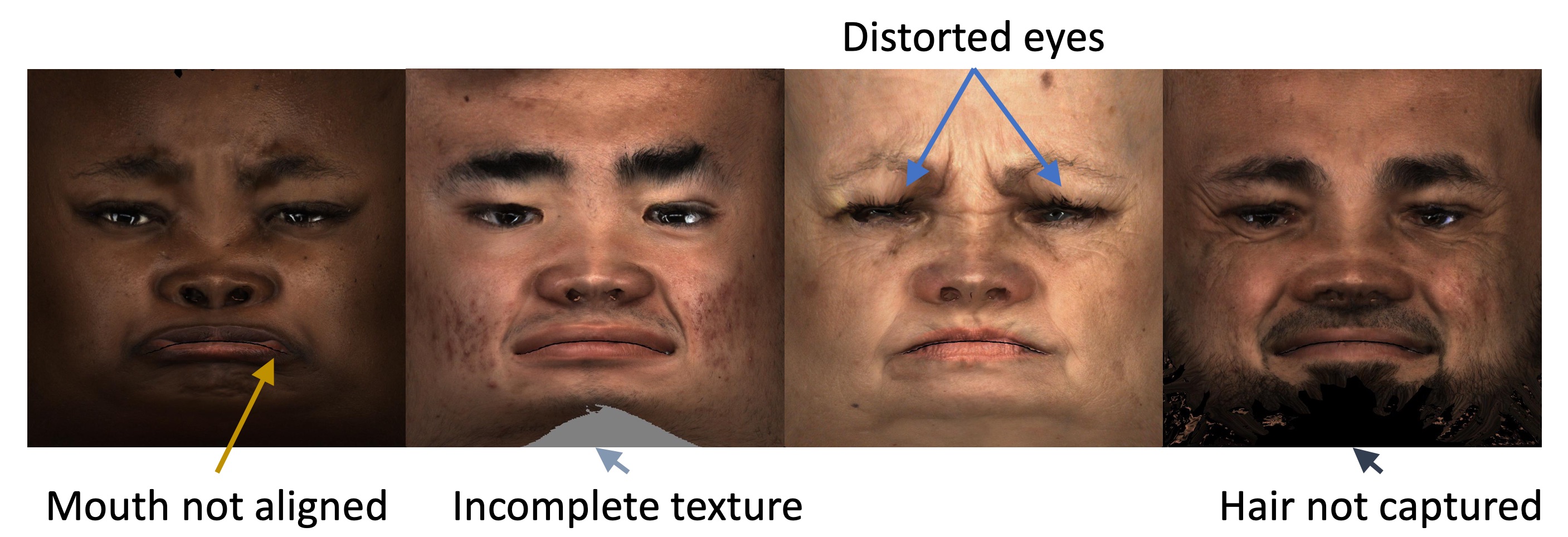}
\caption{Examples of different distortions resulting from the noisy scanning and inexact alignment.}
\label{fig:PDG_corruptions}
\end{figure}

The semi-automatic data acquisition pipeline described in \autoref{sec:training_data_construction} is used to construct a dataset of 2D images that will be used to train the GAN.
Naturally, some of the generated data samples contain corrupted parts due to errors in one or more of the pipeline stages.
In the so-called 3D scanning process, for example, facial textures that contain hair are often not captured well. 
Another reasons for incomplete texture are occlusions and limited camera field of view. 
The geometry of the eyes is occasionally distorted due to their high specular reflection properties. 
In the landmark annotation stage, some landmarks can be inaccurate or even wrong, resulting in various distortions in the final output. \autoref{fig:PDG_corruptions} provides several examples of such data corruptions.
One way to handle data corruption is to ignore imperfect images and keep only the valid ones.
In our case, manual screening of the data reduced the number of samples from $4,963$ to only $3,679$ valid ones, thus, eliminating $25\%$ of the data.
Here, we propose a novel technique for training GANs using partially incomplete data that is able to exploit undamaged parts and robustly deal with corrupted data.

\begin{figure}
\centering
\includegraphics[width=1\linewidth]{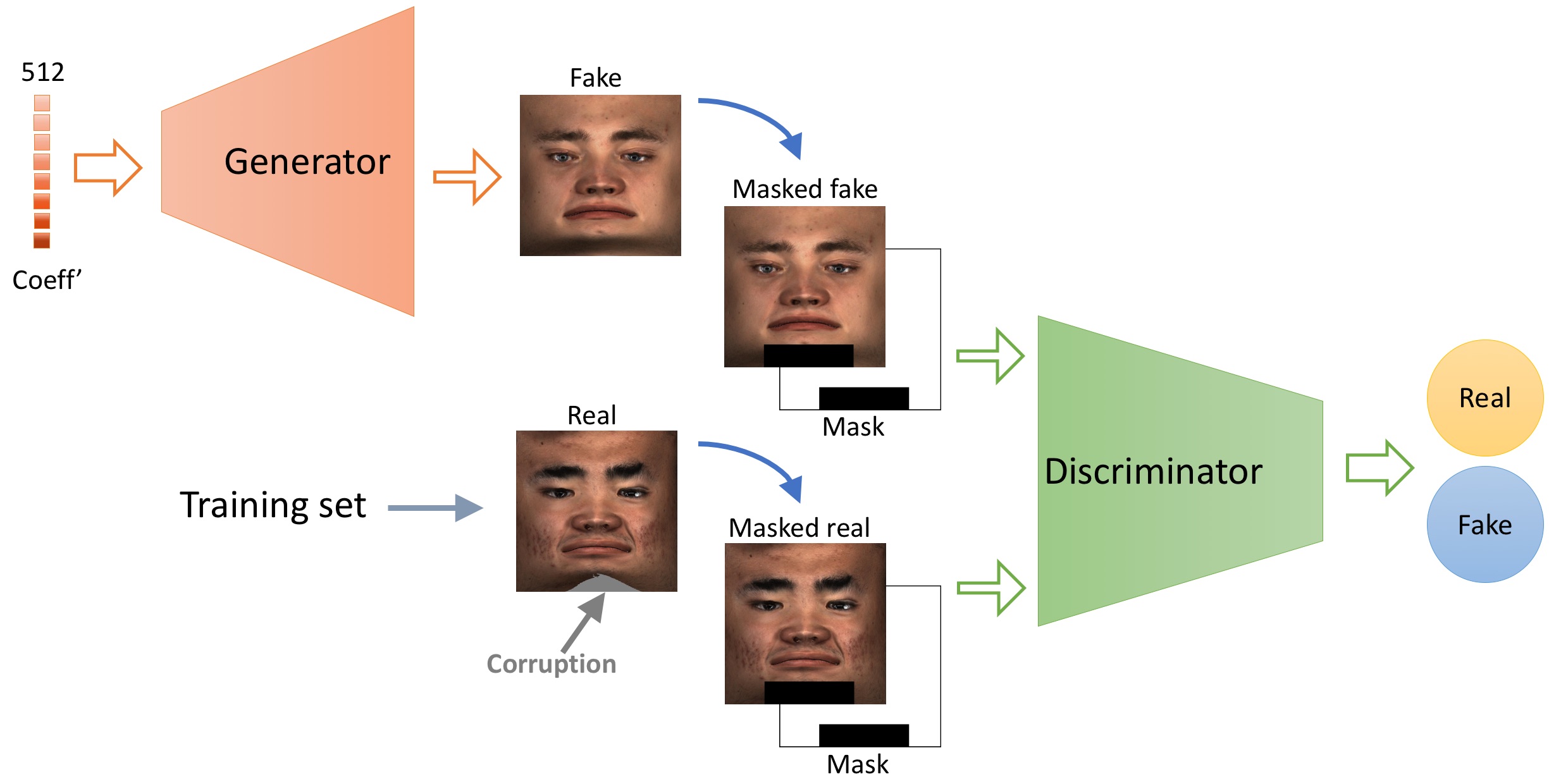}
\caption{The proposed GAN model of learning from incomplete data. 
 A valid mask is fitted for each training sample according to the corrupted regions. 
 The real and fake samples are concatenated and masked by the same valid masks and then passed to the discriminator unit. 
 The discriminator cannot distinguish between the real and fake images by their masks. 
 The generator cannot generate corrupted images since corruptions are masked. 
 The generator does not have any information about the mask and therefore cannot generate corrupted images since they would not fit the masked regions.}
\label{fig:PDG_model}
\end{figure}

To that end, we propose to pair a binary \textit{valid mask} to each training data image, that represents areas in the image that should be ignored. 
Without loss of generality, black areas in the masks (zero values) correspond to corrupted regions in the image we would like to ignore, and white regions (values of one) correspond to valid parts we would like to exploit for training the network.
We propose to multiply these valid masks by their corresponding images, as well as concatenate them as a forth channel (R-G-B-mask).
Recall that the discriminator receives as an input a batch of real images and a batch of fake images.
To prevent the discriminator from discriminating real and fake images by the valid masks, the same masks are multiplied and concatenated to both real and fake batches.
The generator, which does not get the masks as an input, must produce complete images in-painting the masked regions. 
Otherwise, the discriminator would be able to easily identify masked parts that do not match the valid masks and conclude that the image is fake.
The valid masks could be constructed either manually or using automatic image processing technique for detection of the unwanted parts. 
The discriminator and generator of the proposed GAN model are demonstrated in \autoref{fig:PDG_model}.

\begin{figure}
\centering
\includegraphics[width=1\linewidth]{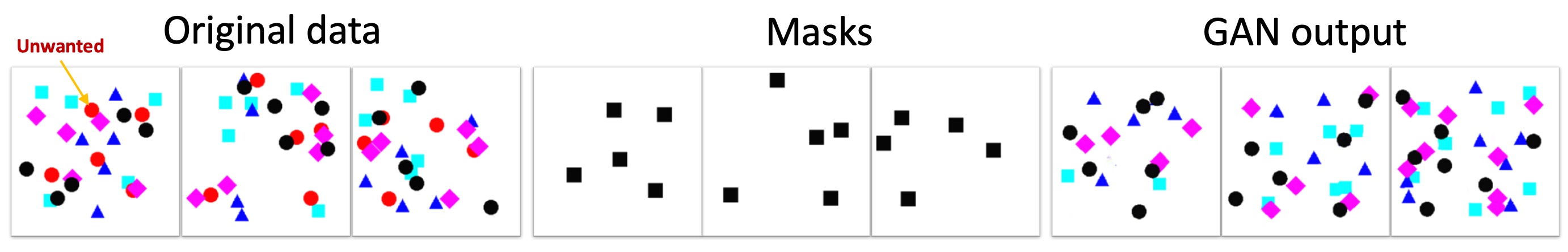}
\caption{Left three: examples of the constructed synthetic data, that consists of images with different colored shapes. 
In this construction, red circles are unwanted in the output. Middle three: valid masks that correspond to the unwanted elements. 
Right three: examples of the output data, generated by the proposed GAN.}
\label{fig:PDG}
\end{figure}

To demonstrate the performance of the proposed GAN we constructed a synthetic dataset of different colored shapes randomly located in $10,000$ images of size $256 \times 256$. 
In this simple experiment, we treat the red circles as corruptions that we would like our model to ignore. \autoref{fig:PDG} shows the data images, the valid masks, and the resulting GAN output. 
It is clearly seen that the proposed GAN model generated new data images without the unwanted red circles.
%
%
\section{Facial Surface Generator}
We propose to train a model which is able to generate realistic geometries and photometries (textures or colors) of human faces. 
The training data for our model is constructed according to \autoref{sec:training_data_construction}, and used to train a NN according to a GAN loss. 
At inference, the trained model is used to produce random plausible facial textures which are mapped by our predefined parametrization described in \autoref{sec:mapping}. 
In order to also generate corresponding facial geometries for each new texture, we propose two novel approaches. 
The first approach is based on training a similar model for geometries. This is done by mapping the training set geometry coordinates using the canonical parametrization into the unit rectangle. By treating each coordinate as a color channel, an we form geometry images which we use to train our geometry generator model.
The second approach relies on the classical 3DMM model. 
For both approaches we suggest a method to generate a geometry which is a plausible fit for a given texture. 
In the following sections we describe the two proposed approaches in detail.
\subsection{Generating textures using GAN}
\begin{figure}
\centering
\includegraphics[width=1\linewidth]{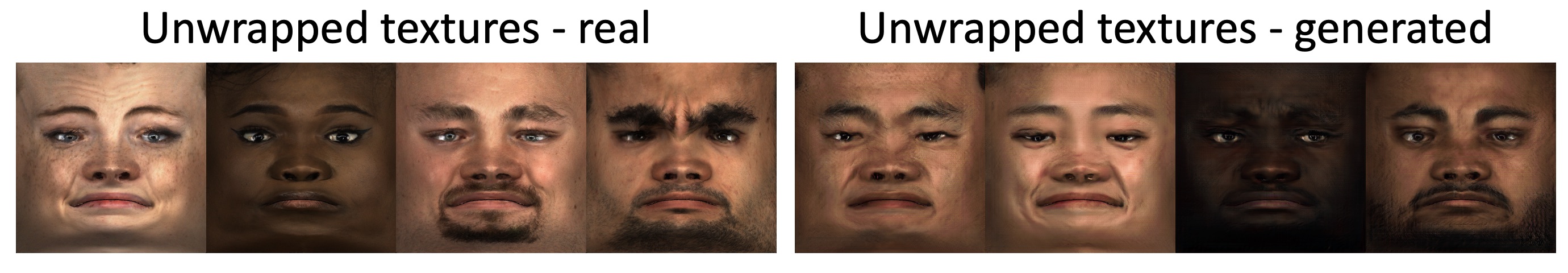}
\caption{Left: Facial textures generated by the suggested pipeline. Right: Real textures from training set.}
\label{fig:fake_real_textures}
\end{figure}

Our texture generation model is based on a Convolutional Neural Network which is trained using a GAN loss. 
Due to this loss, we are able to train a model that satisfies the distribution of real data samples, by drawing new samples out of this distribution. 
By training our model on our dataset which we constructed according to \autoref{sec:training_data_construction}, we are able generate new plausible textures which are all mapped to the unit rectangle plane according to the predefined parametrization described in \autoref{sec:mapping}. 
As we will show in the following sections, the generated textures by the proposed model present novel yet realistic human faces. 
Since texture and geometry are both inseparable attributes of the same geometric entity, it is necessary to take the relationship between them into account when generating the corresponding geometries. 
In \autoref{sec:geom_gen} and \autoref{sec:geom_fit} we describe in detail the process of the proposed geometry generation pipeline which takes as input a generated texture and produces a corresponding plausible geometry. 
Several outputs from the suggested texture generation model are depicted in \autoref{fig:fake_real_textures}

\subsection{Assigning geometries to textures} 
\label{sec:geom_fit}
Once novel textures have been generated, we would like to assign them plausible synthetic geometries in order to obtain realistic face models.
%
One way to generate geometries is by exploiting the 3DMM model by which geometries can be recovered through proper selection of the coefficients.
In what follows, we discuss and compare several methods for obtaining the 3DMM geometry coefficients.

\subsubsection{Random}
The simplest way of synthesizing a geometry to a given texture is by picking random 3DMM geometry coefficients.
We follow the formulation in \autoref{eq:3dmm_dist}. 
The probability of a coefficient $\alpha_i$ is given by
\begin{equation}
P(\alpha_i) \sim \exp\left \{-\frac{\alpha_i^2}{2\sigma_i^2}\right \},
\end{equation}
where, $\sigma_i^2$ is the $i$-th eigenvalue of the covariance of $\Delta G$. 
$\sigma_i^2$ can be computed more efficiently as $\sigma_i^2 = \frac{1}{n}\delta_i^2$, where $\delta_i$ is the $i$-th singular value of $\Delta G$.
To fit a geometry to a given texture, we randomize a vector of coefficients from the above probability distribution and reconstruct the geometry using the 3DMM formulation.

\begin{figure}
\centering
\includegraphics[width=0.4\linewidth]{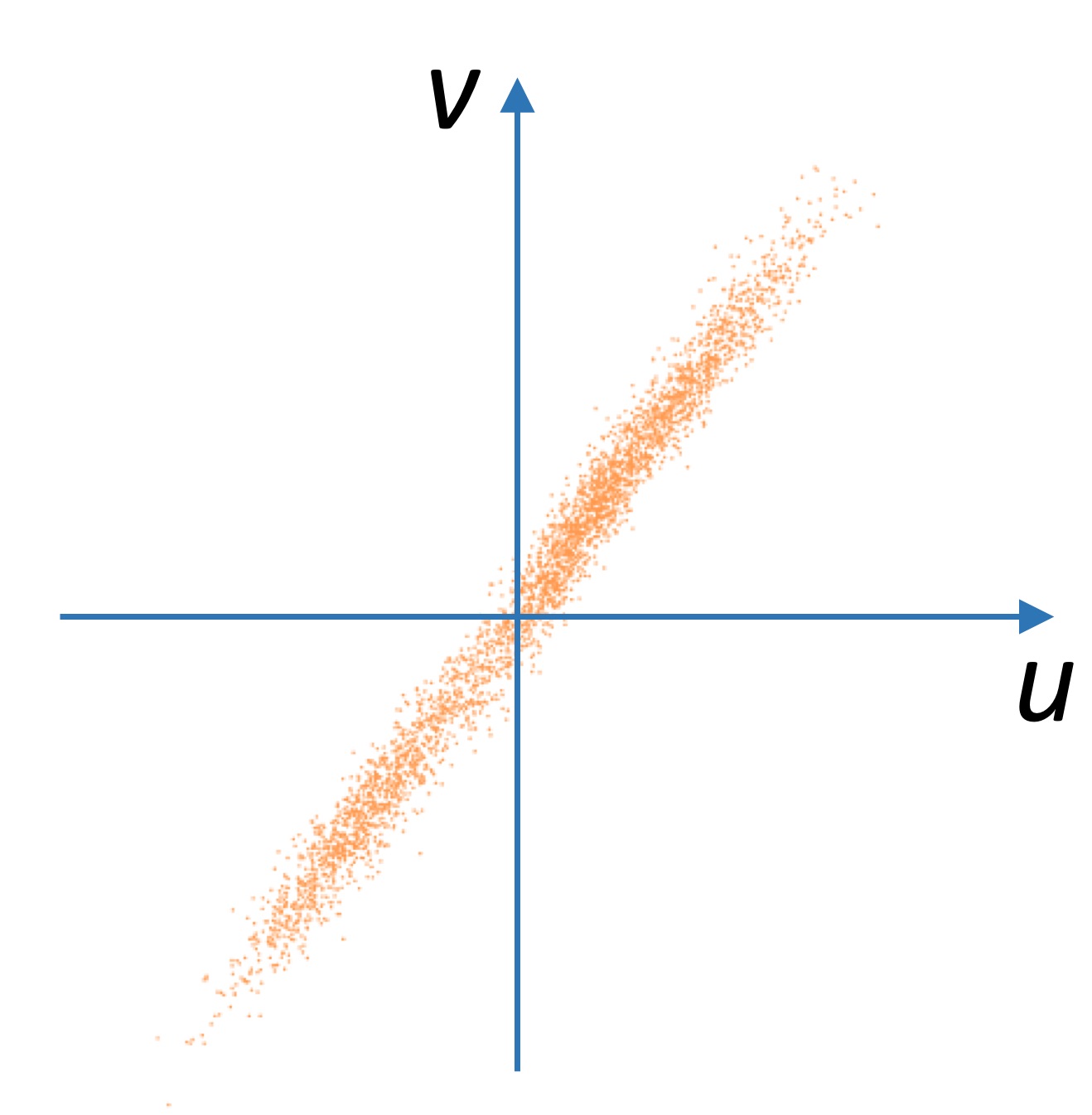}
\caption{
Visualization of the correlation between textures and geometries of the facial scans, where the axes $u$ and $v$ are the first two variables of the canonical correlation.
%
}
\label{fig:fit_corr}
\end{figure}

Random geometries are simple to generate. 
Yet, not every geometry can actually fit any texture.
As a convincing visualization, we computed the \textit{canonical correlation} \cite{hotelling1936relations} between the 3DMM texture and geometry coefficients, $\{\alpha_{ti}\}_{i=1}^n$ and $\{\alpha_{gi}\}_{i=1}^n$, of the facial scans. \autoref{fig:fit_corr} shows $u$ w.r.t. $v$, the first two canonical variables of the correlation.
In what follows, we attempt to generate geometries which are suited for their textures.

\subsubsection{Nearest neighbour}
Given a new texture, a simple way to fit a geometry that is likely to match it, is by finding the data sample with the nearest texture, and projecting its geometry onto the 3DMM subspace.
For that task, we define a distance between two textures as the $L_2$ norm between their 3DMM texture coefficients.
Only the 3DMM texture and geometry coefficients of the data need to be stored.
Nearest neighbor geometries are simple to obtain; however they are restricted to the training data geometries alone.

\subsubsection{Maximum likelihood}
The maximum likelihood estimator (ML) is typically used when one can formulate assumptions about the data distribution. 
In our case, given input facial textures, ML could be used to obtain the most likely geometries under a set of assumptions.
We first construct a mutual 3DMM basis by concatenating textures and geometries.  
Define a vertical concatenation of geometries $G$ and textures $T$ as the $6m \times n$ matrix $M=\begin{pmatrix}G\\T\end{pmatrix}$, 
such that $G,T$ are defined in \autoref{sec:3dmm}.
Define $\Delta M = M - \mu_M\mathbbm{1}^T$, where $\mu_M$ holds the average of the rows of $M$.
Denote by $U$ the $6m \times k$ matrix that contains the first $k$ basis vectors of $\Delta M$, i.e., corresponding to the largest magnitude eigenvalues.
These vectors can be computed either as eigenvectors of $\Delta M \Delta M^T$ or, more efficiently, as the left singular vectors of $\Delta M$.
 Denote by $U_g$ and $\mu_{M_g}$ the upper halves of $U$ and $\mu_{M}$, and denote by $U_t$ and $\mu_{M_t}$ the lower halves $U$ and $\mu_{M}$, respectively, such that
 \begin{equation}
 U =      \begin{pmatrix}
           U_g\\U_t
          \end{pmatrix},\,\,\,\,\,\,\,
\mu_{M} = 
\begin{pmatrix}
\mu_{M_g} \\ \mu_{M_t}
\end{pmatrix}.
 \end{equation}
Note that $U_g$ and $U_t$, unlike $V_g$ and $V_t$ that were defined in \autoref{sec:3dmm}, are not orthogonal. 
Nevertheless, any geometry $g$ and texture $t$ of a given face in $M$ can be represented as a linear combination 
\begin{equation}
    \begin{pmatrix} g \\ t \end{pmatrix} = 
    \begin{pmatrix} \mu_{M_g} \\ \mu_{M_t} \end{pmatrix} + 
    \begin{pmatrix} U_g \\ U_t \end{pmatrix}\beta,
\end{equation}
where the coefficient vector $\beta$ is mutual to the geometry and texture.
Using the notations and definitions above, any new facial texture could be approximated through a coefficient vector $\beta$ as 
\begin{eqnarray}
    t &=& U_t\beta + \mu_{M_t} + noise_t. \cr
    g &=& U_g\beta + \mu_{M_g} + noise_g.
    \label{eq:texture_noise_model}
\end{eqnarray}
The maximum likelihood assumption is that $\beta$, $noise_t$, and $noise_g$ follow a multivariate normal distributions with zero mean.
Given a facial texture $t$, our goal is to compute the most likely coefficient vector $\beta^*$ under this assumption, and then obtain the most likely geometry as 
\begin{equation}
    g = U_g\beta^* + \mu_{M_g}.
\end{equation}
Following Bayes' rule, one could formulate the most likely coefficient vector as
\begin{eqnarray}
\beta^* &=& \argmax_{\beta} P(\beta|t)\cr 
 & =&  \argmax_{\beta} \frac{P(t|\beta)P(\beta)}{P(t)}  \cr
 & =& \argmax_{\beta} P(t|\beta)P(\beta).
\end{eqnarray}
Since $P(t|\beta)$ and $P(\beta)$ follow multivariate normal distributions, denote their covariance matrices by $\Sigma_{t|\beta}$ and $\Sigma_{\beta}$, and their mean vectors by $\mu_{t|\beta} = U_t\beta$ and $\mu_\beta = \vec{0}$, respectively.
Thus,
\begin{eqnarray}
\beta^* &=& \argmax_{\beta} P(t|\beta)P(\beta)  \cr
 &=& \argmax_\beta \exp\left \{-\frac{1}{2}(t-\mu_{t|\beta})\Sigma_{t|\beta}^{-1}(t-\mu_{t|\beta})^T\right \}\cdot
\exp\left \{-\frac{1}{2}\beta\Sigma_{\beta}^{-1}\beta^T\right \}  \cr
&=& \argmin_{\beta} (t-U_t\beta)\Sigma_{t|\beta}^{-1}(t-V_t\beta)^T + \beta\Sigma_{\beta}^{-1}\beta^T \,\,\,. 
\end{eqnarray}
One could obtain a closed form solution for $\beta^*$ by vanishing its gradient, which yields
\begin{equation}
\beta^* = (U_t^T\Sigma_{t|\beta}^{-1}U_t+\Sigma_{\beta})^{-1}(t^T\Sigma_{t|\beta}^{-1}U_t).
\end{equation}
We estimate the covariance matrices $\Sigma_{\beta}$ and $\Sigma_{t|\beta}$ empirically from the data. Since the mean of each coefficient in $\beta$ with respect to all samples is zero, the covariance $\Sigma_{\beta}$ can be estimated by
\begin{equation}
    \Sigma_{\beta} = \frac{1}{n-1}\sum_{i=1}^{n}\beta_i\beta_i^T,
\end{equation}
where, $\beta_i$ is the coefficient vector for face sample $i$.
The $3m \times 3m$ covariance matrix $\Sigma_{t|\beta}$ is very large, impractical to estimate from a few thousands of samples or to invert once estimated. 
Hence, for simplicity, we approximate it as a diagonal matrix that does not depend on $\beta$. 
One can verify that the mean of each element in $noise_t = U_t\beta + \mu_{M_t} - t$ with respect to all samples is zero. 
Hence, we estimate its $j$-th diagonal value as
\begin{equation}
    \Sigma_{t|\beta,jj} = \frac{1}{n-1}\sum_{i=1}^{n}noise_{t,ij}^2.
\end{equation}

\subsubsection{Least squares}
Least squares (LS) minimization is a simple and very useful approach that should be typically used when the amount of data samples is large enough.
It can be thought of as training a multivariate linear regression with an $L_2$ loss. 
Assume a facial sample is represented by a texture vector $t$ and a geometry vector $g$. 
Denote by $\alpha_t$ and $\alpha_g$ the column vectors with the first $k_t$ and $k_g$ texture and geometry 3DMM coefficients of the face. 
These coefficients can be obtained by projecting $t$ and $g$ onto the 3DMM basis $V_t$ and $V_g$.
Let the $k_t \times n$ matrix $A_t$ hold the texture coefficient vectors of all samples in its columns, and let the $k_g \times n$ matrix $A_g$ hold the geometry coefficient vectors of all samples in its columns in the same order as $A_t$.
The correlation between $A_t$ and $A_g$ could be linearly approximated by
\begin{equation}
    A_g \approx W^TA_t.
    \label{eq:LS}
\end{equation}
Note that we would not benefit from generalizing from a linear to an affine correlation. 
This is because the mean of each row in $A_t$ and $A_g$ are zero, as they hold singular values of a centered set of samples.
Following \autoref{eq:LS}, we would like to find a matrix $W$ that minimizes
\begin{equation}
    \mbox{loss}(W) = \|W^TA_t - A_g\|_{F}.
\end{equation}
A closed form solution is easily obtained to be
\begin{equation}
    W^* = (A_tA_t^T)^{-1}A_tA_g^T = A_t^{+}A_g^T.
\end{equation}
Define $\tilde V_t$ and $\tilde V_g$ as holding the first $k_t$ and $k_g$ texture and geometry 3DMM basis vectors.
$W^*$ can be estimated using a set of training samples. 
Then, given a new texture $t$, one could fit a geometry $g$ by computing the texture coefficients as
\begin{equation}
\alpha_t = \tilde V_t^T(t - \mu_t),
\end{equation}
computing the geometry coefficients as
\begin{equation}
\alpha_g=W^*\alpha_t,
\end{equation}
and finally, computing the geometry as
\begin{equation}
g = \tilde V_g \alpha_g + \mu_g.
\end{equation}

\subsubsection{Geometry reconstruction method comparison} 
\begin{figure}
\centering
\includegraphics[width=1\linewidth]{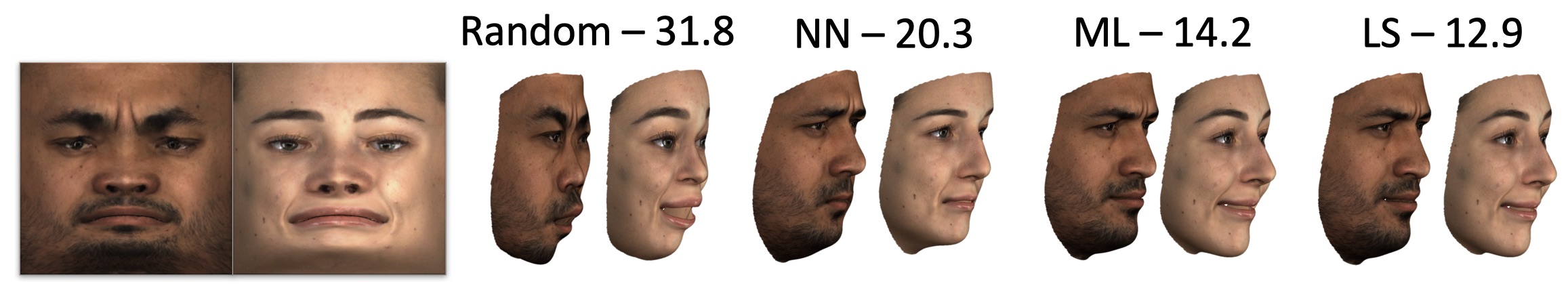}
\caption{ Left: Mapped textures from the test set. Right: Corresponding geometries that were reconstructed using the each proposed approach. 
Average L2 reconstruction error for each method appears above each method illustration.}
\label{fig:geom_fit_1}
\end{figure}

To evaluate how well the assigned geometries fit each textures, we use a test set of textures and their corresponding geometries obtained from $297$ scans that did not participate in neither the GAN training or geometry fitting procedures. 
Given these unseen textures as input, we estimate their geometries using the approaches presented above, and compare them to their ground truth. For this comparison, we computed the average $L_2$ norm between the vertices of the reconstructed and true geometry for each of the methods.
In this experiment, we chose $k = k_t = k_g = 200$.
\autoref{fig:geom_fit_1} shows examples of test textures mapped onto their assigned geometries that were obtained using each of the above methods.

\begin{figure}
\centering
\includegraphics[width=1\linewidth]{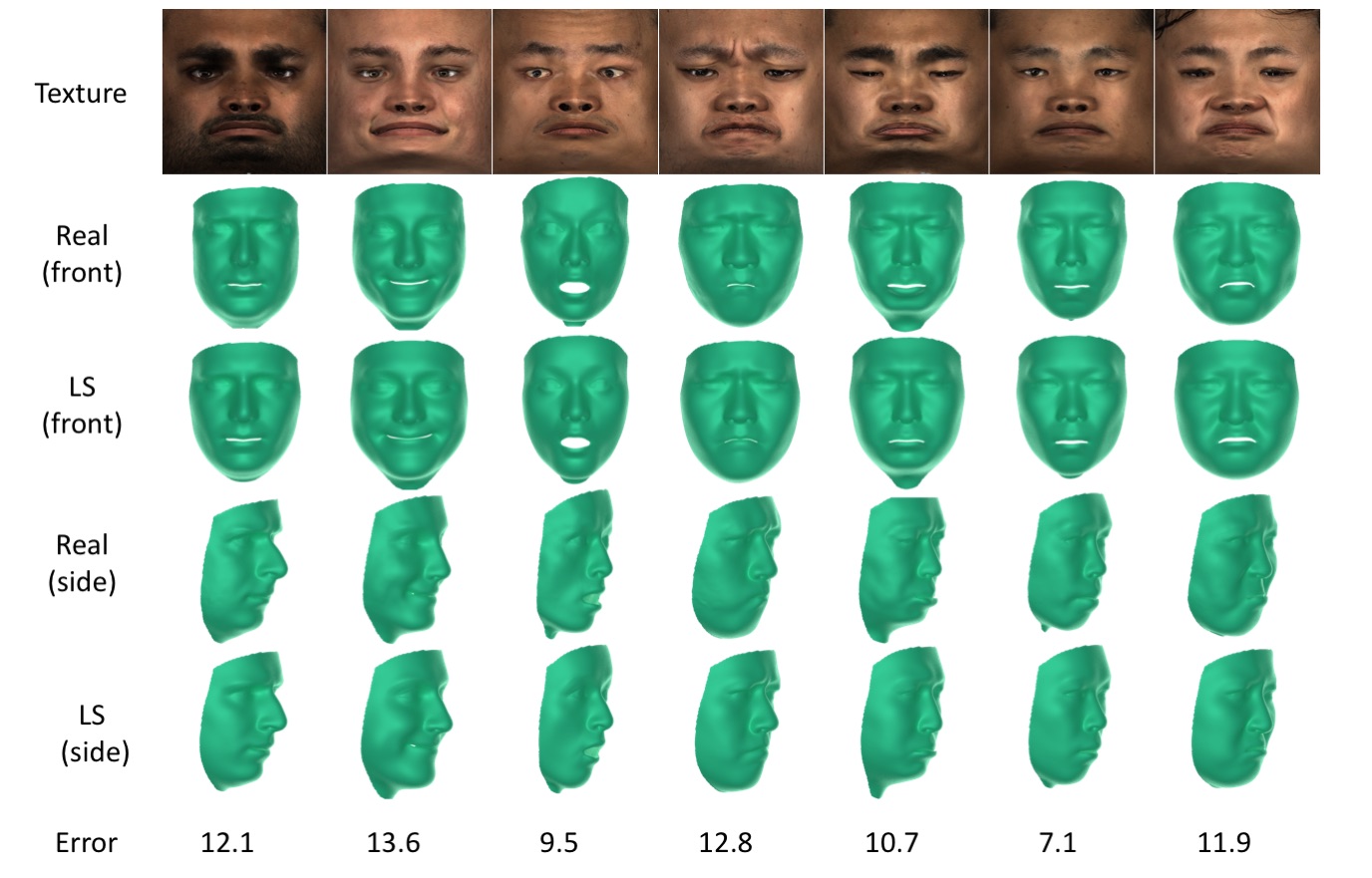}
\caption{Top: examples of unwrapped textures of test scans. 
Below each texture we show the corresponding scanned geometry, and compare it to the one estimated using the least squares approach. 
The $L_2$ norm between the reconstruction and true geometries is indicated at the bottom.}
\label{fig:geom_fit_2}
\end{figure}

It is clear that the LS approach obtains the best results on the test set. 
Since it is also simple and efficient, we choose to use the LS approach to approximate the geometries in the following sections.
Note, however, that the rest of the methods could be beneficial for other applications, depending on each case. \autoref{fig:geom_fit_2} visually compares the reconstructed geometries to the true ones for different textures from the test scans, using the LS approach.
The $L_2$ norm between the reconstruction and true geometries are given below each example. 
The geometries, predicted solely from textures of identities that were never seen before, are surprisingly very similar to the true ones. 
This validates the strong correlation assumption between textures and geometries.

\section{Generating geometries using GAN} 
\label{sec:geom_gen}
In \autoref{sec:geom_fit}, we reconstructed geometries using the 3DMM model. 
Indeed, projecting geometries onto the subspace of 3DMM has almost no visual effect on the appearance of the faces.
The 3DMM, however, is constrained to the subspace of training set geometries and cannot generalize to unseen examples.

In \autoref{sec:mapping}, we mapped facial textures into 2D images, with the goal of producing new textures.
The same methodology can be used for producing new geometries as well.
To that end, we propose to construct a dataset of aligned facial geometries and train a GAN to generate new ones by repeating the texture mapping process while replacing its RBG texture values by its XYZ geometry values.

\begin{figure}
\centering
\includegraphics[width=1\linewidth]{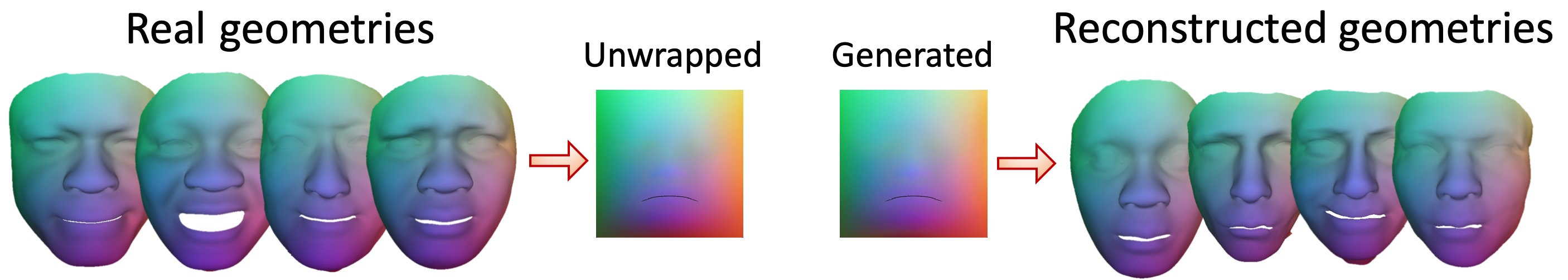}
\caption{Left: the template aligned to $4$ real 3D facial scans, colored by their $X$, $Y$, and $Z$ values. Center-left: geometry values mapped to an image using the proposed universal mapping. These images are used to train a GAN.
Center-right: fake geometry image generated by the GAN.
Right: the synthesized geometries mapped back to 3D.
}
\label{fig:generating_geom}
\end{figure}

As for data augmentation, while the amount of texture samples can be doubled by horizontal mirroring each of the images, we found that mirroring each one of the $X$, $Y$, and $Z$ values independently results in a valid facial geometry. Thus, the amount of geometry samples can be augmented by a factor of $8$.
Note, however, that when mirroring $X$ values, one should perform $C - X$, where $C$ is a constant that could be set, for example, to the maximal $X$ value in all training samples.
The training data and resulting generated geometries are shown in \autoref{fig:generating_geom}.

\section{Adding expressions}
\label{sec:expressions}
In previous sections we proposed a method for generating new facial textures and corresponding geometries. 
In order to complete the model we must also take expressions into consideration. 
We follow \cite{chu20143d} who define a linear basis for expressions by taking the difference $g_{diff} = g_{expr}-g_{neutral}$ for every face in the set. We then remove the mean difference vector to obtain $\Delta G_{diff} = G_{diff}-\mu_{diff}\mathbbm{1}^T$. We compute the principal components of $\Delta G_{diff}$ to obtain our geometric expression model. The expression difference model can be used by randomizing the expression coefficients and adding the linear combination of the difference vectors to a generated neutral face as so $g_{exp} =  g_{neutral} + \mu_{diff} + \alpha_{g} + V_{exp} \alpha_{exp}$.

Since the expression model must be applied to neutral faces, we should  define a model for neutral faces. 
In order to span only the space of neutral expressionless faces we suggest to replace all the geometries in our training set with their neutral counterpart. 
By following this course, the texture model still benefits from all the textures available to us while our geometry model learns to predict only neutral models for any texture with or without expression. 
This method can be applied to either the 3DMM based geometry model from \autoref{sec:geom_fit} or the GAN based geometry model described in \autoref{sec:geom_gen}.
\begin{figure}
\centering
\includegraphics[width=1\linewidth]{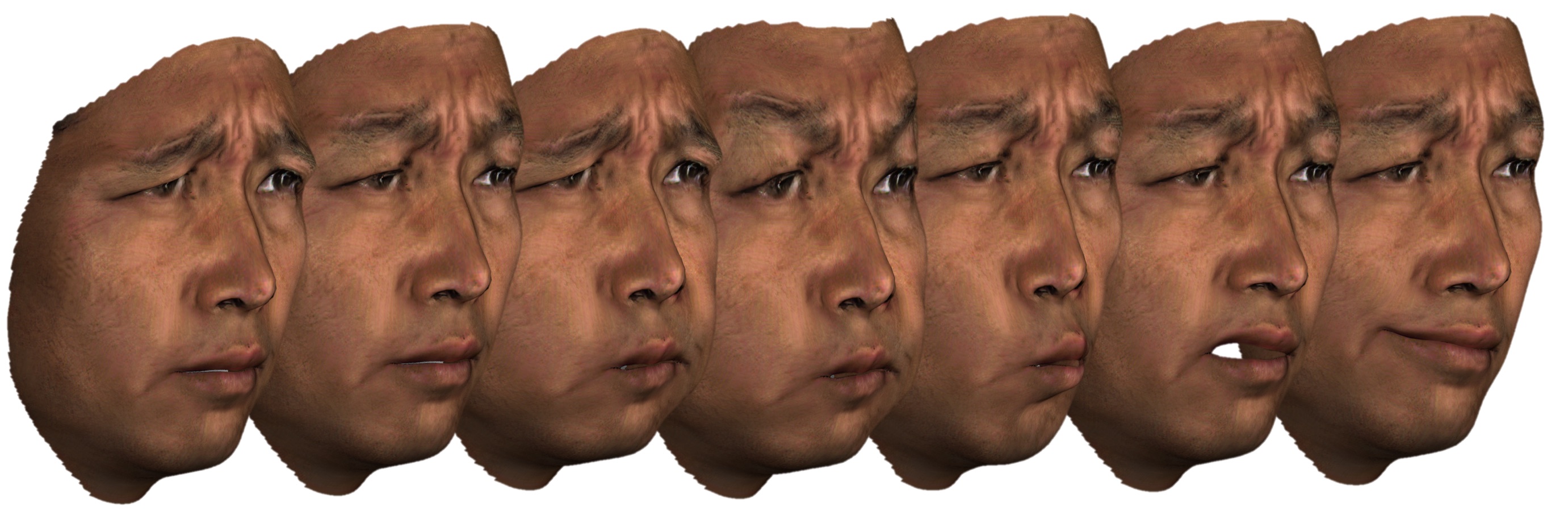}
\caption{Expression vectors applied to a single face.}
\label{fig:blend_expression}
\end{figure}

\section{Experimental results}
\label{sec:expermintal}
\begin{figure}
\centering
\includegraphics[width=1\linewidth]{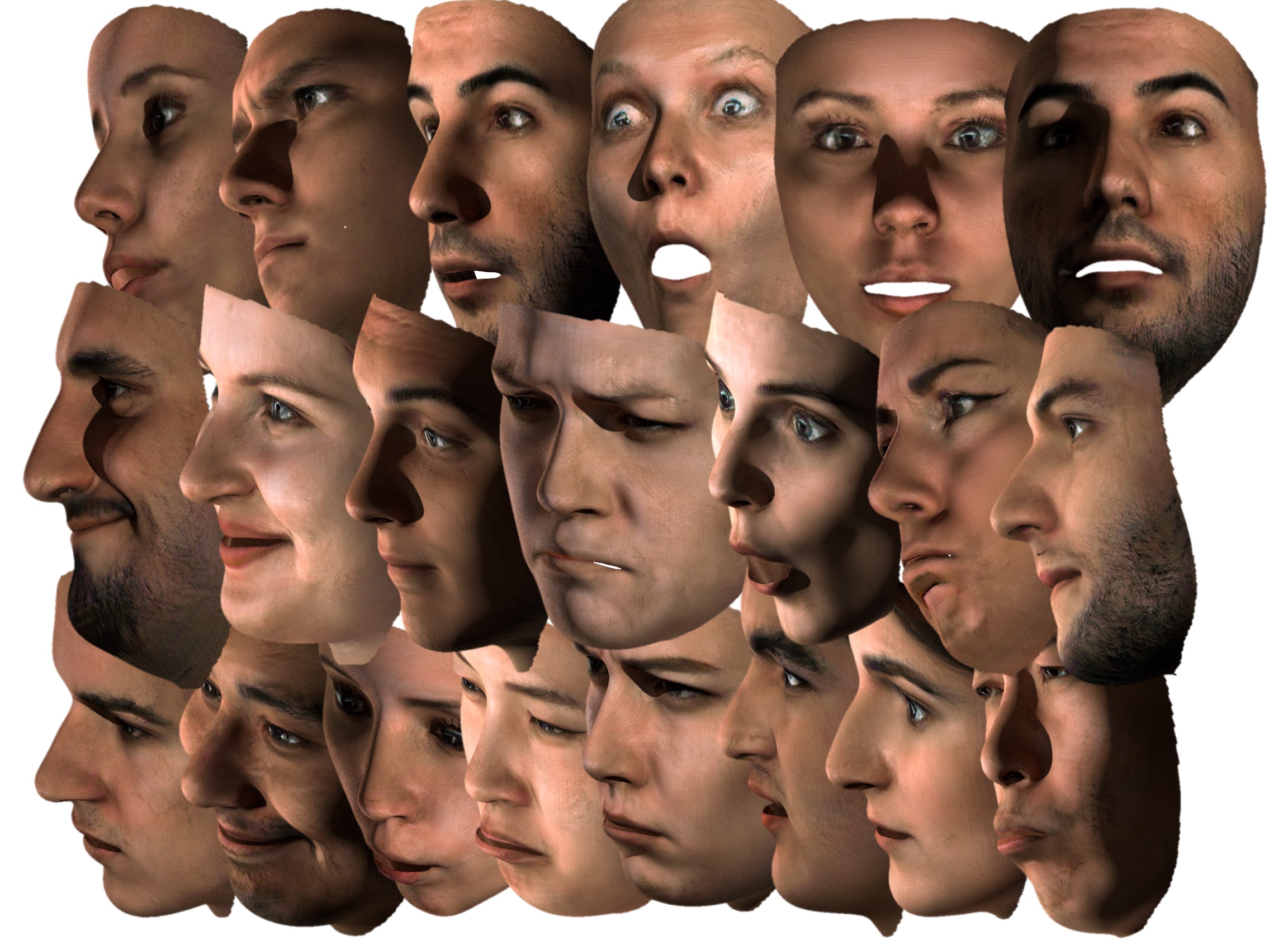}
\caption{Several generated faces rendered with varying expression and pose under varying lighting conditions.}
\label{fig:render_experiment}
\end{figure}

In order to demonstrate the ability of our model to generate new realistic identities we perform several quantitative as well as qualitative experiments. As our first results we generate several random textures and obtain their corresponding geometry according to \autoref{sec:geom_fit}. We are able to vary the expression by applying a linear expression model as described in \autoref{sec:expressions}. According to this model each expression can be represented by a variation from the mean face which leads to a specific facial movement. By combining various movements of the face one can generate any expression desired. The faces are then rendered under various poses and lighting conditions. The rendered faces are depicted in \autoref{fig:render_experiment}

\begin{figure}
\centering
\includegraphics[width=0.8\linewidth]{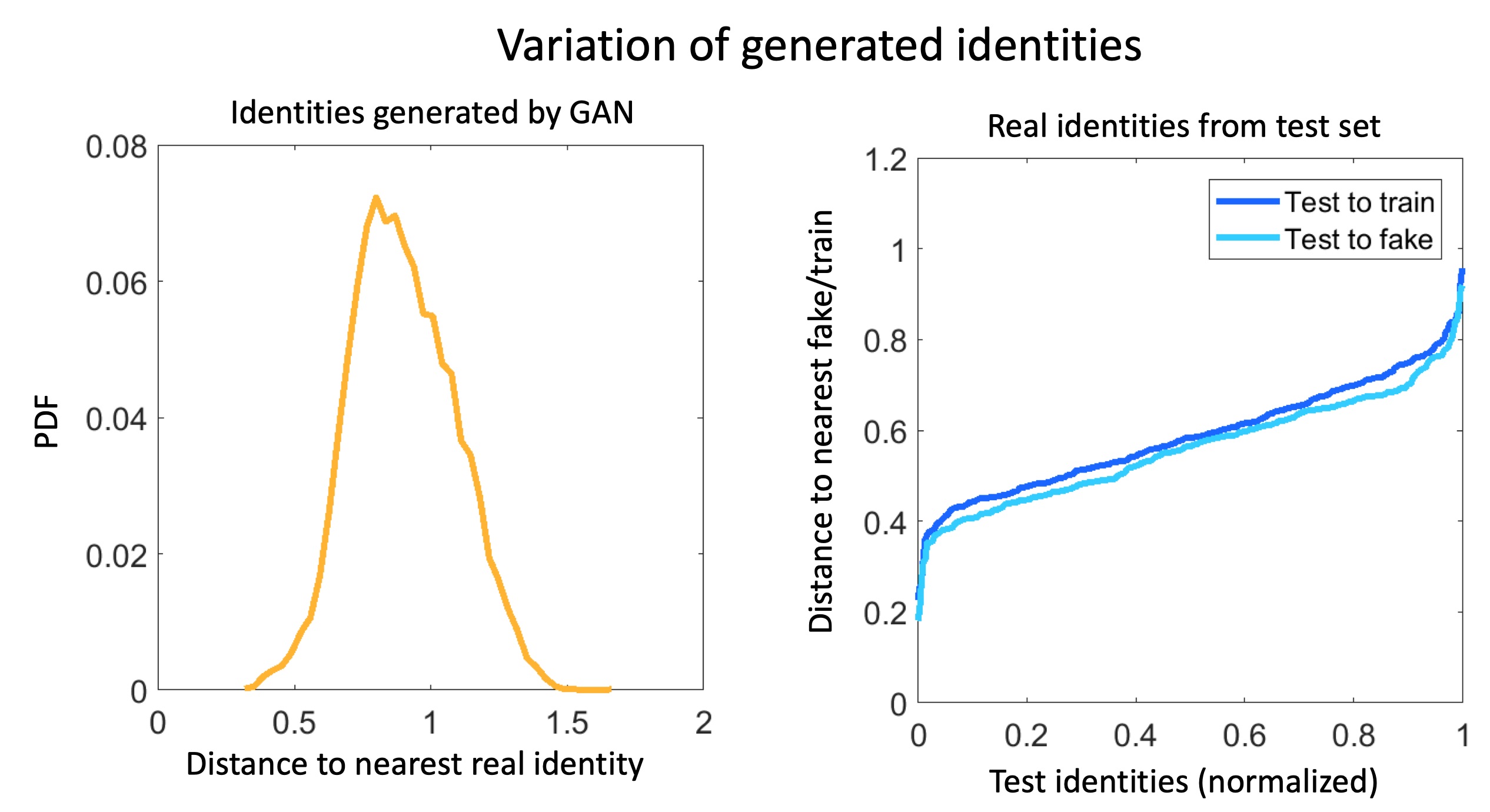}
\caption{Left: Distribution of distances between training and generated IDs. Right: Cumulative sum of distances between training and test in light dark blue and generated to test in light blue.}
\label{fig:ID_var}
\end{figure}

\begin{figure}
\centering
\includegraphics[width=1\linewidth]{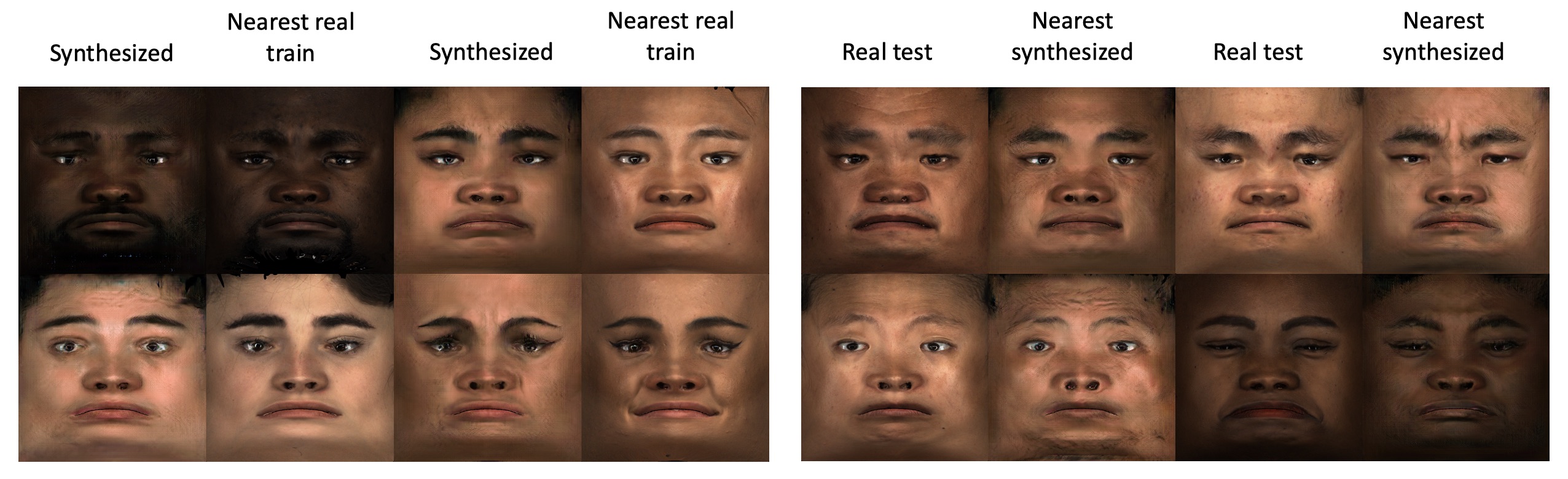}
\caption{Left: Generated textures coupled with their nearest neighbor within the training set. Right: Validation textures coupled with the nearest out of 10k generated texture. }
\label{fig:NN-exp}
\end{figure}

Our next qualitative experiment demonstrates the ability of our model to generate completely new textures by combining facial features from different training examples. To this end we search for the nearest neighbor to the generated texture from within the training data. It can be seen in \autoref{fig:ID_var} that the demonstrated examples have nearest neighbors that are significantly different from them and cannot be considered as the same identity. Within the following section we will analyze both the generative ability of our model to produce new faces as well its realism and similarity to realistic examples. In addition, we also search for generated texture samples which are nearest to several validation set examples. By finding close by textures we demonstrate the generalization capability of our model to unseen textures. This is demonstrated in \autoref{fig:NN-exp}

The previous qualitative assessment is complimented by a more in depth examination of the nearest neighbors across $10K$ generated faces. For the following experiments we have freedom to choose our distance metric. We aim to find a natural metric which coincides with human perception of faces. We therefore choose to render each generated and real face from a frontal perspective and process the rendered images via a pre-trained facial recognition network. Using a model based on \cite{amos2016openface} we extract the last feature from within the network as our facial descriptor. The distance is calculated as $dist=\|D_1 - D_2\|_{2}$ where $D_1,D_2$ are the descriptors corresponding to the first and second face respectively. By analyzing the distribution of such distances we can assess the spread of identities which exists within each dataset as well as the relation between different datasets.

We use the distribution of distances between generated faces and the training and validation sets of real faces in order to assess the quality of our generative model. In \autoref{fig:ID_var} we plot the distribution of distances between generated sample and their nearest real training sample. This plot implies that these distances are distributed as a shifted Gaussian. This implies that on average new identities are located in between the real samples and not directly near them. Our analysis of the distances to the neighbors of the validation set also depicted in \autoref{fig:ID_var} shows that our model is able to produce identities of subjects similar to ones found in the validation set which were not used to train the model. This validates our claim that our model is producing new identities by combining features from the training set faces, and that these identities are not too similar to the training set yet can still generalize to the unseen validation set.

\begin{figure}
\centering
\includegraphics[width=1\linewidth]{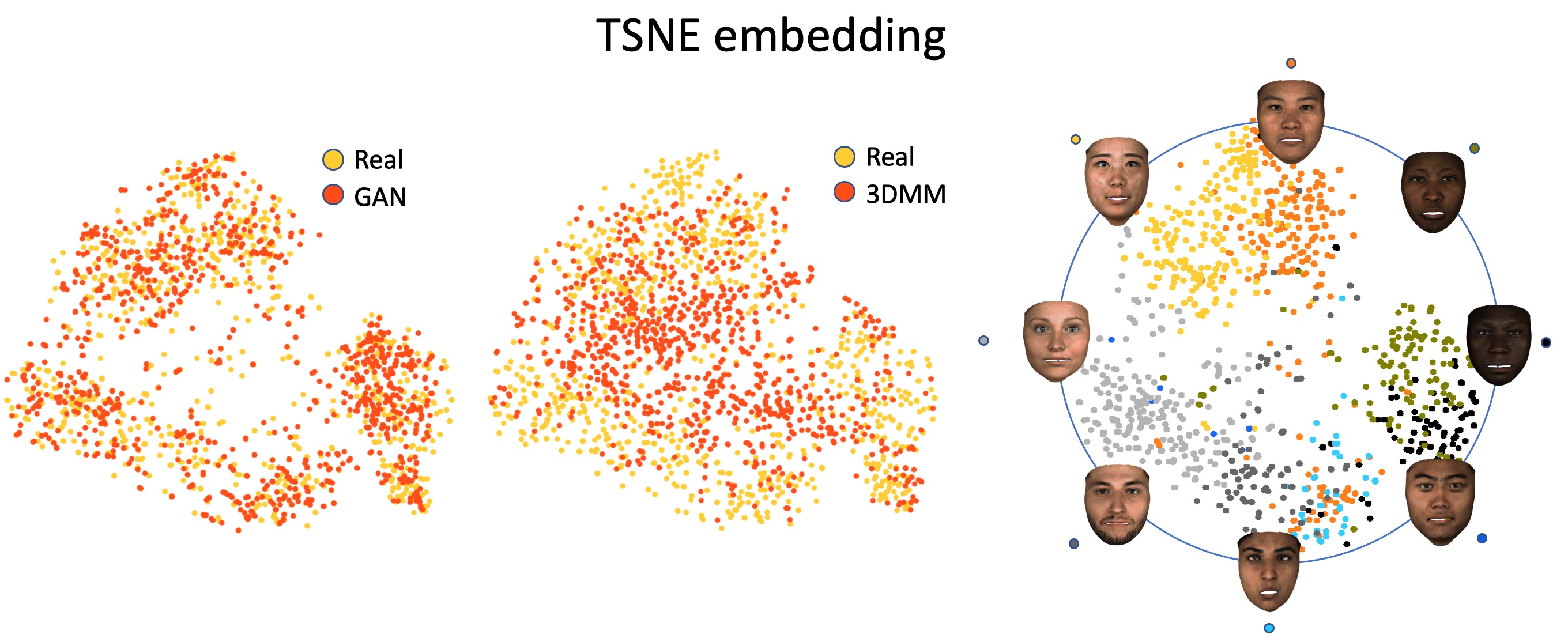}
\caption{T-SNE embedding of face ID's. Left: Real versus generated ID's. Center: Real versus 3DMM ID's Right: generated ID's labeled according to real data clusters}
\label{fig:T-SNE-exp}
\end{figure}

Following \cite{karras2017progressive} we perform an analysis of sliced Wasserstein distance \cite{rabin2011wasserstein} on our generated textures and geometries. By assessing the distance between the distribution of patches taken from textures generated by our model relative to patches taken from faces generated by 3DMM we can analyze the multi-resolution similarity between the generated and real examples. \autoref{tbl:SWD_texture} and \autoref{tbl:SWD_geom} show the SWD for each resolution relative to the training data. In both experiments it is clear that the SWD is lower for our model at every resolution indicating that at every level of detail the patches produced by our model are more similar to patches in the training data. 

In addition to assessing the patch level feature resemblance, we wish to uncover the distances between the distribution of identities. To this end we conduct two more experiments which gauge the similarity between the distributions of generated identities to that of the real ones. In order to qualitatively assess these distributions we depict our identities using the common dimensionality reduction scheme T-SNE \cite{maaten2008visualizing}. \autoref{fig:T-SNE-exp} depicts the low dimensional representation of the embedding proposed by our model and the 3DMM overlaid on top of the real data embedding. In addition \autoref{fig:T-SNE-exp} also depicts the clustering of different ethnic groups as well as gender as data points of different colors. By assigning each generated sample to the nearest cluster, we can automatically assign each new sample with its nearest cluster in order to obtain automatic annotation of our generated data. In addition we perform a quantitative analysis of the difference between identity distribution using SWD. The results of this experiment are depicted in \autoref{tbl:SWD_ids}.

\begin{table}[]
\centering
{\small
\begin{tabularx}{\linewidth}{lXXXXXXXX}
Resolution & 1024 & 512 &  256&  128&  64& 32& 16& avg\\ \hline
Real & 3.33 & 3.33 & 3.35 & 2.93 & 2.53 & 2.47 & 4.16 & 3.16\\
\textbf{{Proposed}} &  35.32 & 18.13 & 10.76 & 6.41 & 7.42 & 10.76 & 34.86 & 17.67\\
PCA &92.7 & 224.01 & 156.87 & 66.20 & 15.71 & 33.97 & 104.08 & 99.08\\
\end{tabularx}
}
\caption{Sliced Wasserstein distance between generated and real texture images.  }
\label{tbl:SWD_texture}

{\small
\begin{tabularx}{\linewidth}{lXXXXXXXX}
Resolution & 1024 & 512 &  256&  128&  64& 32& 16& avg\\ \hline
Real & 6.08 & 2.41 & 3.40 & 2.45 & 3.10 & 2.75 & 1.86 & 3.15\\
\textbf{{Proposed}} & 11.8 & 9.58 & 27.13 & 44.5 & 38.05 & 11.03 & 2.16 & 20.61\\
PCA & 272.7 & 43.94 & 29.6 & 50.72 & 44.43 & 13.21 & 4.67 & 65.61 \\
\end{tabularx}
}
\caption{Sliced Wasserstein distance between generated and real geometry images. }
\label{tbl:SWD_geom}

{\small
\begin{tabularx}{\linewidth}{lXXXX}
Method & 3DMM & \textbf{{Proposed}} & train\\ \hline
train & 59.88 & 35.82 & - \\
test & 75.3 & 62.09 & 42.4 \\
\end{tabularx}
}
\caption{Sliced Wasserstein distance between distributions of identities from different sets.  }
\label{tbl:SWD_ids}

\end{table}
\section{Discussion}
In this paper we present a new model for generating high detail textures and corresponding geometries of human faces. Our claim is that an effective method for processing geometric surfaces via CNNs is to first align the geometric dataset to a template model, and then map each geometry to a 2D image using a predefined mapping. Once in image form, A GAN loss can be employed to train a generator model which aims to imitate the distribution of the training data images. We further show that by training a generator for both textures and geometries it is possible to synthesize high-detail textures and geometries which are linked to each other by a single canonical mapping. 

In addition, we describe in \autoref{sec:geom_fit} several methods for fitting 3DMM geometries by learning the underlying relation between texture and geometry, a relation that has been largely neglected in previous work. In \autoref{sec:geom_fit} we also provide a quantitative and qualitative evaluation of each geometry reconstruction method. our proposed face generation pipeline therefore consists of a high resolution texture generator combined with a geometry that was either produced by a similar geometric generation model or by employing a learning scheme which produces the most likely corresponding 3DMM coefficients.

Besides the main pipeline, we propose two extra data processing steps which improve sample quality. In \autoref{sec:mapping} we describe the design and construction of our canonical mapping. Our mapping by design is intended to reduce distortion in important high detail areas while spreading the flattening distortion to non essential areas. Our mapping was also designed in order to take maximal advantage of the available area in each image. In \autoref{sec:mapping} we also show that our improved mapping compared to \cite{slossberg2018high} indeed preserves delicate texture details in our predefined high importance regions. In \autoref{sec:corrupted_data} we also present a new technique for dealing with partially corrupted data. This is especially important when the data acquisition is expensive and prone to errors. By adding a corruption mask to the data at train time the network is able to ignore the affected areas while still learning from the mostly unaffected ones. In the case of our dataset this increases the amount of usable data by roughly $20\%$.

In order to evaluate our proposed model we preformed a quantitative as well as qualitative analysis of several aspects of our model. Our main objective was to create a realistic model, a requirement which we break down into several factors. Our model should produce high quality plausible facial textures which look as much like the training data as possible, but also compose new faces not seen during training rather than repeat previously seen faces. To that end we use an efficient approximation of Wasserstein distance between distributions in order to evaluate the local and global scale features of the produced textures and geometries as well as the distance between distributions of real and generated identities. Our results show that in both identity distribution and image feature resemblance we outperform the 3DMM model which the most widely used model to date.

\begin{acks}

This research was partially supported by the Israel Ministry of Science, grant number  3-14719 and the Technion Hiroshi Fujiwara Cyber Security Research Center and the Israel Cyber Bureau.
We would also like to thank Intel RealSense group for sharing their data and computational resources with us. 

\end{acks}

\bibliographystyle{ACM-Reference-Format}
\bibliography{refs}


\begin{thebibliography}{37}


\ifx \showCODEN    \undefined \def \showCODEN     #1{\unskip}     \fi
\ifx \showDOI      \undefined \def \showDOI       #1{#1}\fi
\ifx \showISBNx    \undefined \def \showISBNx     #1{\unskip}     \fi
\ifx \showISBNxiii \undefined \def \showISBNxiii  #1{\unskip}     \fi
\ifx \showISSN     \undefined \def \showISSN      #1{\unskip}     \fi
\ifx \showLCCN     \undefined \def \showLCCN      #1{\unskip}     \fi
\ifx \shownote     \undefined \def \shownote      #1{#1}          \fi
\ifx \showarticletitle \undefined \def \showarticletitle #1{#1}   \fi
\ifx \showURL      \undefined \def \showURL       {\relax}        \fi
\providecommand\bibfield[2]{#2}
\providecommand\bibinfo[2]{#2}
\providecommand\natexlab[1]{#1}
\providecommand\showeprint[2][]{arXiv:#2}

\bibitem[\protect\citeauthoryear{Amos, Ludwiczuk, and Satyanarayanan}{Amos
  et~al\mbox{.}}{2016}]%
        {amos2016openface}
\bibfield{author}{\bibinfo{person}{Brandon Amos}, \bibinfo{person}{Bartosz
  Ludwiczuk}, {and} \bibinfo{person}{Mahadev Satyanarayanan}.}
  \bibinfo{year}{2016}\natexlab{}.
\newblock \bibinfo{booktitle}{\emph{OpenFace: A general-purpose face
  recognition library with mobile applications}}.
\newblock \bibinfo{type}{{T}echnical {R}eport}.
  \bibinfo{institution}{CMU-CS-16-118, CMU School of Computer Science}.
\newblock


\bibitem[\protect\citeauthoryear{Bas, Huber, Smith, Awais, and Kittler}{Bas
  et~al\mbox{.}}{2017}]%
        {bas20173d}
\bibfield{author}{\bibinfo{person}{Anil Bas}, \bibinfo{person}{Patrik Huber},
  \bibinfo{person}{William~AP Smith}, \bibinfo{person}{Muhammad Awais}, {and}
  \bibinfo{person}{Josef Kittler}.} \bibinfo{year}{2017}\natexlab{}.
\newblock \showarticletitle{3d morphable models as spatial transformer
  networks}. In \bibinfo{booktitle}{\emph{Proc. ICCV Workshop on Geometry Meets
  Deep Learning}}. \bibinfo{pages}{904--912}.
\newblock


\bibitem[\protect\citeauthoryear{Blanz and Vetter}{Blanz and Vetter}{1999}]%
        {blanz1999morphable}
\bibfield{author}{\bibinfo{person}{Volker Blanz} {and} \bibinfo{person}{Thomas
  Vetter}.} \bibinfo{year}{1999}\natexlab{}.
\newblock \showarticletitle{A morphable model for the synthesis of 3D faces}.
  In \bibinfo{booktitle}{\emph{Proceedings of the 26th annual conference on
  Computer graphics and interactive techniques}}. ACM Press/Addison-Wesley
  Publishing Co., \bibinfo{pages}{187--194}.
\newblock


\bibitem[\protect\citeauthoryear{{Blender Online Community}}{{Blender Online
  Community}}{2017}]%
        {blender}
\bibfield{author}{\bibinfo{person}{{Blender Online Community}}.}
  \bibinfo{year}{2017}\natexlab{}.
\newblock \bibinfo{booktitle}{\emph{Blender - a 3D modelling and rendering
  package}}.
\newblock Blender Foundation, Blender Institute, Amsterdam.
\newblock
\newblock
\shownote{\url{http://www.blender.org}.}


\bibitem[\protect\citeauthoryear{Booth, Antonakos, Ploumpis, Trigeorgis,
  Panagakis, Zafeiriou, et~al\mbox{.}}{Booth et~al\mbox{.}}{2017}]%
        {booth20173d}
\bibfield{author}{\bibinfo{person}{James Booth}, \bibinfo{person}{Epameinondas
  Antonakos}, \bibinfo{person}{Stylianos Ploumpis}, \bibinfo{person}{George
  Trigeorgis}, \bibinfo{person}{Yannis Panagakis}, \bibinfo{person}{Stefanos
  Zafeiriou}, {et~al\mbox{.}}} \bibinfo{year}{2017}\natexlab{}.
\newblock \showarticletitle{3D Face Morphable Models “In-the-Wild”}. In
  \bibinfo{booktitle}{\emph{Proceedings of the IEEE Conference on
  ComputerVision and Pattern Recognition}}.
\newblock


\bibitem[\protect\citeauthoryear{Booth, Roussos, Ponniah, Dunaway, and
  Zafeiriou}{Booth et~al\mbox{.}}{2018}]%
        {booth2018large}
\bibfield{author}{\bibinfo{person}{James Booth}, \bibinfo{person}{Anastasios
  Roussos}, \bibinfo{person}{Allan Ponniah}, \bibinfo{person}{David Dunaway},
  {and} \bibinfo{person}{Stefanos Zafeiriou}.} \bibinfo{year}{2018}\natexlab{}.
\newblock \showarticletitle{Large scale 3d morphable models}.
\newblock \bibinfo{journal}{\emph{International Journal of Computer Vision}}
  \bibinfo{volume}{126}, \bibinfo{number}{2-4} (\bibinfo{year}{2018}),
  \bibinfo{pages}{233--254}.
\newblock


\bibitem[\protect\citeauthoryear{Booth, Roussos, Zafeiriou, Ponniah, and
  Dunaway}{Booth et~al\mbox{.}}{2016}]%
        {booth20163d}
\bibfield{author}{\bibinfo{person}{James Booth}, \bibinfo{person}{Anastasios
  Roussos}, \bibinfo{person}{Stefanos Zafeiriou}, \bibinfo{person}{Allan
  Ponniah}, {and} \bibinfo{person}{David Dunaway}.}
  \bibinfo{year}{2016}\natexlab{}.
\newblock \showarticletitle{A 3d morphable model learnt from 10,000 faces}. In
  \bibinfo{booktitle}{\emph{Proceedings of the IEEE Conference on Computer
  Vision and Pattern Recognition}}. \bibinfo{pages}{5543--5552}.
\newblock


\bibitem[\protect\citeauthoryear{Bronstein, Bruna, LeCun, Szlam, and
  Vandergheynst}{Bronstein et~al\mbox{.}}{2017}]%
        {bronstein2017geometric}
\bibfield{author}{\bibinfo{person}{Michael~M Bronstein}, \bibinfo{person}{Joan
  Bruna}, \bibinfo{person}{Yann LeCun}, \bibinfo{person}{Arthur Szlam}, {and}
  \bibinfo{person}{Pierre Vandergheynst}.} \bibinfo{year}{2017}\natexlab{}.
\newblock \showarticletitle{Geometric deep learning: going beyond euclidean
  data}.
\newblock \bibinfo{journal}{\emph{IEEE Signal Processing Magazine}}
  \bibinfo{volume}{34}, \bibinfo{number}{4} (\bibinfo{year}{2017}),
  \bibinfo{pages}{18--42}.
\newblock


\bibitem[\protect\citeauthoryear{Chu, Romdhani, and Chen}{Chu
  et~al\mbox{.}}{2014}]%
        {chu20143d}
\bibfield{author}{\bibinfo{person}{Baptiste Chu}, \bibinfo{person}{Sami
  Romdhani}, {and} \bibinfo{person}{Liming Chen}.}
  \bibinfo{year}{2014}\natexlab{}.
\newblock \showarticletitle{3D-aided face recognition robust to expression and
  pose variations}. In \bibinfo{booktitle}{\emph{Proceedings of the IEEE
  Conference on Computer Vision and Pattern Recognition}}.
  \bibinfo{pages}{1899--1906}.
\newblock


\bibitem[\protect\citeauthoryear{Deng, Cheng, Xue, Zhou, and Zafeiriou}{Deng
  et~al\mbox{.}}{2018}]%
        {Deng_2018_CVPR}
\bibfield{author}{\bibinfo{person}{Jiankang Deng}, \bibinfo{person}{Shiyang
  Cheng}, \bibinfo{person}{Niannan Xue}, \bibinfo{person}{Yuxiang Zhou}, {and}
  \bibinfo{person}{Stefanos Zafeiriou}.} \bibinfo{year}{2018}\natexlab{}.
\newblock \showarticletitle{UV-GAN: Adversarial Facial UV Map Completion for
  Pose-Invariant Face Recognition}. In \bibinfo{booktitle}{\emph{The IEEE
  Conference on Computer Vision and Pattern Recognition (CVPR)}}.
\newblock


\bibitem[\protect\citeauthoryear{Floater}{Floater}{1997}]%
        {floater1997parametrization}
\bibfield{author}{\bibinfo{person}{Michael~S Floater}.}
  \bibinfo{year}{1997}\natexlab{}.
\newblock \showarticletitle{Parametrization and smooth approximation of surface
  triangulations}.
\newblock \bibinfo{journal}{\emph{Computer aided geometric design}}
  \bibinfo{volume}{14}, \bibinfo{number}{3} (\bibinfo{year}{1997}),
  \bibinfo{pages}{231--250}.
\newblock


\bibitem[\protect\citeauthoryear{Gecer, Bhattarai, Kittler, and Kim}{Gecer
  et~al\mbox{.}}{2018}]%
        {Gecer_2018_ECCV}
\bibfield{author}{\bibinfo{person}{Baris Gecer}, \bibinfo{person}{Binod
  Bhattarai}, \bibinfo{person}{Josef Kittler}, {and} \bibinfo{person}{Tae-Kyun
  Kim}.} \bibinfo{year}{2018}\natexlab{}.
\newblock \showarticletitle{Semi-supervised Adversarial Learning to Generate
  Photorealistic Face Images of New Identities from 3D Morphable Model}. In
  \bibinfo{booktitle}{\emph{The European Conference on Computer Vision
  (ECCV)}}.
\newblock


\bibitem[\protect\citeauthoryear{Goodfellow, Pouget-Abadie, Mirza, Xu,
  Warde-Farley, Ozair, Courville, and Bengio}{Goodfellow et~al\mbox{.}}{2014}]%
        {goodfellow2014generative}
\bibfield{author}{\bibinfo{person}{Ian Goodfellow}, \bibinfo{person}{Jean
  Pouget-Abadie}, \bibinfo{person}{Mehdi Mirza}, \bibinfo{person}{Bing Xu},
  \bibinfo{person}{David Warde-Farley}, \bibinfo{person}{Sherjil Ozair},
  \bibinfo{person}{Aaron Courville}, {and} \bibinfo{person}{Yoshua Bengio}.}
  \bibinfo{year}{2014}\natexlab{}.
\newblock \showarticletitle{Generative adversarial nets}. In
  \bibinfo{booktitle}{\emph{Advances in neural information processing
  systems}}. \bibinfo{pages}{2672--2680}.
\newblock


\bibitem[\protect\citeauthoryear{Hotelling}{Hotelling}{1936}]%
        {hotelling1936relations}
\bibfield{author}{\bibinfo{person}{Harold Hotelling}.}
  \bibinfo{year}{1936}\natexlab{}.
\newblock \showarticletitle{Relations between two sets of variates}.
\newblock \bibinfo{journal}{\emph{Biometrika}} \bibinfo{volume}{28},
  \bibinfo{number}{3/4} (\bibinfo{year}{1936}), \bibinfo{pages}{321--377}.
\newblock


\bibitem[\protect\citeauthoryear{Huang, Zhang, Li, and He}{Huang
  et~al\mbox{.}}{2017}]%
        {huang2017beyond}
\bibfield{author}{\bibinfo{person}{Rui Huang}, \bibinfo{person}{Shu Zhang},
  \bibinfo{person}{Tianyu Li}, {and} \bibinfo{person}{Ran He}.}
  \bibinfo{year}{2017}\natexlab{}.
\newblock \showarticletitle{Beyond Face Rotation: Global and Local Perception
  GAN for Photorealistic and Identity Preserving Frontal View Synthesis}. In
  \bibinfo{booktitle}{\emph{Proceedings of the IEEE International Conference on
  Computer Vision}}. \bibinfo{pages}{2439--2448}.
\newblock


\bibitem[\protect\citeauthoryear{Jolliffe}{Jolliffe}{1986}]%
        {jolliffe1986principal}
\bibfield{author}{\bibinfo{person}{Ian~T Jolliffe}.}
  \bibinfo{year}{1986}\natexlab{}.
\newblock \showarticletitle{Principal component analysis and factor analysis}.
\newblock In \bibinfo{booktitle}{\emph{Principal component analysis}}.
  \bibinfo{publisher}{Springer}, \bibinfo{pages}{115--128}.
\newblock


\bibitem[\protect\citeauthoryear{Karras, Aila, Laine, and Lehtinen}{Karras
  et~al\mbox{.}}{2017}]%
        {karras2017progressive}
\bibfield{author}{\bibinfo{person}{Tero Karras}, \bibinfo{person}{Timo Aila},
  \bibinfo{person}{Samuli Laine}, {and} \bibinfo{person}{Jaakko Lehtinen}.}
  \bibinfo{year}{2017}\natexlab{}.
\newblock \showarticletitle{Progressive growing of gans for improved quality,
  stability, and variation}.
\newblock \bibinfo{journal}{\emph{nternational Conference on Learning
  Representations (ICLR)}} (\bibinfo{year}{2017}).
\newblock


\bibitem[\protect\citeauthoryear{King}{King}{2009}]%
        {dlib09}
\bibfield{author}{\bibinfo{person}{Davis~E. King}.}
  \bibinfo{year}{2009}\natexlab{}.
\newblock \showarticletitle{Dlib-ml: A Machine Learning Toolkit}.
\newblock \bibinfo{journal}{\emph{Journal of Machine Learning Research}}
  \bibinfo{volume}{10} (\bibinfo{year}{2009}), \bibinfo{pages}{1755--1758}.
\newblock


\bibitem[\protect\citeauthoryear{Litany, Bronstein, Bronstein, and
  Makadia}{Litany et~al\mbox{.}}{2018}]%
        {Litany_2018_CVPR}
\bibfield{author}{\bibinfo{person}{Or Litany}, \bibinfo{person}{Alex
  Bronstein}, \bibinfo{person}{Michael Bronstein}, {and}
  \bibinfo{person}{Ameesh Makadia}.} \bibinfo{year}{2018}\natexlab{}.
\newblock \showarticletitle{Deformable Shape Completion With Graph
  Convolutional Autoencoders}. In \bibinfo{booktitle}{\emph{The IEEE Conference
  on Computer Vision and Pattern Recognition (CVPR)}}.
\newblock


\bibitem[\protect\citeauthoryear{Maaten and Hinton}{Maaten and Hinton}{2008}]%
        {maaten2008visualizing}
\bibfield{author}{\bibinfo{person}{Laurens van~der Maaten} {and}
  \bibinfo{person}{Geoffrey Hinton}.} \bibinfo{year}{2008}\natexlab{}.
\newblock \showarticletitle{Visualizing data using t-SNE}.
\newblock \bibinfo{journal}{\emph{Journal of machine learning research}}
  \bibinfo{volume}{9}, \bibinfo{number}{Nov} (\bibinfo{year}{2008}),
  \bibinfo{pages}{2579--2605}.
\newblock


\bibitem[\protect\citeauthoryear{Masi, Trần, Hassner, Leksut, and
  Medioni}{Masi et~al\mbox{.}}{2016}]%
        {masi2016we}
\bibfield{author}{\bibinfo{person}{Iacopo Masi}, \bibinfo{person}{Anh~Tuấn
  Trần}, \bibinfo{person}{Tal Hassner}, \bibinfo{person}{Jatuporn~Toy
  Leksut}, {and} \bibinfo{person}{G{\'e}rard Medioni}.}
  \bibinfo{year}{2016}\natexlab{}.
\newblock \showarticletitle{Do we really need to collect millions of faces for
  effective face recognition?}. In \bibinfo{booktitle}{\emph{European
  Conference on Computer Vision}}. Springer, \bibinfo{pages}{579--596}.
\newblock


\bibitem[\protect\citeauthoryear{Rabin, Peyr{\'e}, Delon, and Bernot}{Rabin
  et~al\mbox{.}}{2011}]%
        {rabin2011wasserstein}
\bibfield{author}{\bibinfo{person}{Julien Rabin}, \bibinfo{person}{Gabriel
  Peyr{\'e}}, \bibinfo{person}{Julie Delon}, {and} \bibinfo{person}{Marc
  Bernot}.} \bibinfo{year}{2011}\natexlab{}.
\newblock \showarticletitle{Wasserstein barycenter and its application to
  texture mixing}. In \bibinfo{booktitle}{\emph{International Conference on
  Scale Space and Variational Methods in Computer Vision}}. Springer,
  \bibinfo{pages}{435--446}.
\newblock


\bibitem[\protect\citeauthoryear{Ranjan, Bolkart, Sanyal, and Black}{Ranjan
  et~al\mbox{.}}{2018}]%
        {ranjan2018generating}
\bibfield{author}{\bibinfo{person}{Anurag Ranjan}, \bibinfo{person}{Timo
  Bolkart}, \bibinfo{person}{Soubhik Sanyal}, {and} \bibinfo{person}{Michael~J
  Black}.} \bibinfo{year}{2018}\natexlab{}.
\newblock \showarticletitle{Generating 3D Faces Using Convolutional Mesh
  Autoencoders}. In \bibinfo{booktitle}{\emph{European Conference on Computer
  Vision}}. Springer, \bibinfo{pages}{725--741}.
\newblock


\bibitem[\protect\citeauthoryear{Richardson, Sela, and Kimmel}{Richardson
  et~al\mbox{.}}{2016}]%
        {richardson20163d}
\bibfield{author}{\bibinfo{person}{Elad Richardson}, \bibinfo{person}{Matan
  Sela}, {and} \bibinfo{person}{Ron Kimmel}.} \bibinfo{year}{2016}\natexlab{}.
\newblock \showarticletitle{3D face reconstruction by learning from synthetic
  data}. In \bibinfo{booktitle}{\emph{3D Vision (3DV), 2016 Fourth
  International Conference on}}. IEEE, \bibinfo{pages}{460--469}.
\newblock


\bibitem[\protect\citeauthoryear{Richardson, Sela, Or-El, and
  Kimmel}{Richardson et~al\mbox{.}}{2017}]%
        {richardson2017learning}
\bibfield{author}{\bibinfo{person}{Elad Richardson}, \bibinfo{person}{Matan
  Sela}, \bibinfo{person}{Roy Or-El}, {and} \bibinfo{person}{Ron Kimmel}.}
  \bibinfo{year}{2017}\natexlab{}.
\newblock \showarticletitle{Learning detailed face reconstruction from a single
  image}. In \bibinfo{booktitle}{\emph{2017 IEEE Conference on Computer Vision
  and Pattern Recognition (CVPR)}}. IEEE, \bibinfo{pages}{5553--5562}.
\newblock


\bibitem[\protect\citeauthoryear{Riza Alp~G\"uler}{Riza Alp~G\"uler}{2016}]%
        {Guler2016DenseReg}
\bibfield{author}{\bibinfo{person}{Epameinondas Antonakos Patrick Snape
  Stefanos Zafeiriou Iasonas~Kokkinos Riza Alp~G\"uler, George~Trigeorgis}.}
  \bibinfo{year}{2016}\natexlab{}.
\newblock \showarticletitle{DenseReg: Fully Convolutional Dense Shape
  Regression In-the-Wild}.
\newblock \bibinfo{journal}{\emph{arXiv:1612.01202}} (\bibinfo{year}{2016}).
\newblock


\bibitem[\protect\citeauthoryear{Saito, Wei, Hu, Nagano, and Li}{Saito
  et~al\mbox{.}}{2017}]%
        {saito2017photorealistic}
\bibfield{author}{\bibinfo{person}{Shunsuke Saito}, \bibinfo{person}{Lingyu
  Wei}, \bibinfo{person}{Liwen Hu}, \bibinfo{person}{Koki Nagano}, {and}
  \bibinfo{person}{Hao Li}.} \bibinfo{year}{2017}\natexlab{}.
\newblock \showarticletitle{Photorealistic facial texture inference using deep
  neural networks}. In \bibinfo{booktitle}{\emph{IEEE Conference on Computer
  Vision and Pattern Recognition, CVPR}}, Vol.~\bibinfo{volume}{3}.
\newblock


\bibitem[\protect\citeauthoryear{Sela, Richardson, and Kimmel}{Sela
  et~al\mbox{.}}{2017}]%
        {sela2017unrestricted}
\bibfield{author}{\bibinfo{person}{Matan Sela}, \bibinfo{person}{Elad
  Richardson}, {and} \bibinfo{person}{Ron Kimmel}.}
  \bibinfo{year}{2017}\natexlab{}.
\newblock \showarticletitle{Unrestricted facial geometry reconstruction using
  image-to-image translation}. In \bibinfo{booktitle}{\emph{2017 IEEE
  International Conference on Computer Vision (ICCV)}}. IEEE,
  \bibinfo{pages}{1585--1594}.
\newblock


\bibitem[\protect\citeauthoryear{Shamai, Zibulevsky, and Kimmel}{Shamai
  et~al\mbox{.}}{2018}]%
        {shamai2018efficient}
\bibfield{author}{\bibinfo{person}{Gil Shamai}, \bibinfo{person}{Michael
  Zibulevsky}, {and} \bibinfo{person}{Ron Kimmel}.}
  \bibinfo{year}{2018}\natexlab{}.
\newblock \showarticletitle{Efficient Inter-Geodesic Distance Computation and
  Fast Classical Scaling}.
\newblock \bibinfo{journal}{\emph{IEEE transactions on pattern analysis and
  machine intelligence}} (\bibinfo{year}{2018}).
\newblock


\bibitem[\protect\citeauthoryear{Shrivastava, Pfister, Tuzel, Susskind, Wang,
  and Webb}{Shrivastava et~al\mbox{.}}{2017}]%
        {shrivastava2017learning}
\bibfield{author}{\bibinfo{person}{Ashish Shrivastava}, \bibinfo{person}{Tomas
  Pfister}, \bibinfo{person}{Oncel Tuzel}, \bibinfo{person}{Josh Susskind},
  \bibinfo{person}{Wenda Wang}, {and} \bibinfo{person}{Russ Webb}.}
  \bibinfo{year}{2017}\natexlab{}.
\newblock \showarticletitle{Learning from simulated and unsupervised images
  through adversarial training}. In \bibinfo{booktitle}{\emph{The IEEE
  Conference on Computer Vision and Pattern Recognition (CVPR)}},
  Vol.~\bibinfo{volume}{3}. \bibinfo{pages}{6}.
\newblock


\bibitem[\protect\citeauthoryear{Slossberg, Shamai, and Kimmel}{Slossberg
  et~al\mbox{.}}{2018}]%
        {slossberg2018high}
\bibfield{author}{\bibinfo{person}{Ron Slossberg}, \bibinfo{person}{Gil
  Shamai}, {and} \bibinfo{person}{Ron Kimmel}.}
  \bibinfo{year}{2018}\natexlab{}.
\newblock \showarticletitle{High Quality Facial Surface and Texture Synthesis
  via Generative Adversarial Networks}.
\newblock \bibinfo{journal}{\emph{arXiv preprint arXiv:1808.08281}}
  (\bibinfo{year}{2018}).
\newblock


\bibitem[\protect\citeauthoryear{Tewari, Zollhöfer, Garrido, Bernard, Kim,
  Pérez, and Theobalt}{Tewari et~al\mbox{.}}{2018}]%
        {Tewari_2018_CVPR}
\bibfield{author}{\bibinfo{person}{Ayush Tewari}, \bibinfo{person}{Michael
  Zollhöfer}, \bibinfo{person}{Pablo Garrido}, \bibinfo{person}{Florian
  Bernard}, \bibinfo{person}{Hyeongwoo Kim}, \bibinfo{person}{Patrick Pérez},
  {and} \bibinfo{person}{Christian Theobalt}.} \bibinfo{year}{2018}\natexlab{}.
\newblock \showarticletitle{Self-Supervised Multi-Level Face Model Learning for
  Monocular Reconstruction at Over 250 Hz}. In \bibinfo{booktitle}{\emph{The
  IEEE Conference on Computer Vision and Pattern Recognition (CVPR)}}.
\newblock


\bibitem[\protect\citeauthoryear{Tran, Hassner, Masi, and Medioni}{Tran
  et~al\mbox{.}}{2017a}]%
        {tran2017regressing}
\bibfield{author}{\bibinfo{person}{Anh~Tuan Tran}, \bibinfo{person}{Tal
  Hassner}, \bibinfo{person}{Iacopo Masi}, {and} \bibinfo{person}{G{\'e}rard
  Medioni}.} \bibinfo{year}{2017}\natexlab{a}.
\newblock \showarticletitle{Regressing robust and discriminative 3D morphable
  models with a very deep neural network}. In \bibinfo{booktitle}{\emph{2017
  IEEE Conference on Computer Vision and Pattern Recognition (CVPR)}}. IEEE,
  \bibinfo{pages}{1493--1502}.
\newblock


\bibitem[\protect\citeauthoryear{Tran and Liu}{Tran and Liu}{2018}]%
        {tran2018nonlinear}
\bibfield{author}{\bibinfo{person}{Luan Tran} {and} \bibinfo{person}{Xiaoming
  Liu}.} \bibinfo{year}{2018}\natexlab{}.
\newblock \showarticletitle{Nonlinear 3D Face Morphable Model}.
\newblock \bibinfo{journal}{\emph{arXiv preprint arXiv:1804.03786}}
  (\bibinfo{year}{2018}).
\newblock


\bibitem[\protect\citeauthoryear{Tran, Yin, and Liu}{Tran
  et~al\mbox{.}}{2017b}]%
        {tran2017disentangled}
\bibfield{author}{\bibinfo{person}{Luan Tran}, \bibinfo{person}{Xi Yin}, {and}
  \bibinfo{person}{Xiaoming Liu}.} \bibinfo{year}{2017}\natexlab{b}.
\newblock \showarticletitle{Disentangled Representation Learning GAN for
  Pose-Invariant Face Recognition}. In \bibinfo{booktitle}{\emph{Proceedings of
  the IEEE Conference on Computer Vision and Pattern Recognition}}.
  \bibinfo{pages}{1415--1424}.
\newblock


\bibitem[\protect\citeauthoryear{Weise, Li, Van~Gool, and Pauly}{Weise
  et~al\mbox{.}}{2009}]%
        {weise2009face}
\bibfield{author}{\bibinfo{person}{Thibaut Weise}, \bibinfo{person}{Hao Li},
  \bibinfo{person}{Luc Van~Gool}, {and} \bibinfo{person}{Mark Pauly}.}
  \bibinfo{year}{2009}\natexlab{}.
\newblock \showarticletitle{Face/off: Live facial puppetry}. In
  \bibinfo{booktitle}{\emph{Proceedings of the 2009 ACM SIGGRAPH/Eurographics
  Symposium on Computer animation}}. ACM, \bibinfo{pages}{7--16}.
\newblock


\bibitem[\protect\citeauthoryear{Wu, Zhang, Xue, Freeman, and Tenenbaum}{Wu
  et~al\mbox{.}}{2016}]%
        {wu2016learning}
\bibfield{author}{\bibinfo{person}{Jiajun Wu}, \bibinfo{person}{Chengkai
  Zhang}, \bibinfo{person}{Tianfan Xue}, \bibinfo{person}{Bill Freeman}, {and}
  \bibinfo{person}{Josh Tenenbaum}.} \bibinfo{year}{2016}\natexlab{}.
\newblock \showarticletitle{Learning a probabilistic latent space of object
  shapes via 3d generative-adversarial modeling}. In
  \bibinfo{booktitle}{\emph{Advances in Neural Information Processing
  Systems}}. \bibinfo{pages}{82--90}.
\newblock


\end{thebibliography}

\end{document}